\documentclass[11pt,titlepage]{article}%
\usepackage{amsfonts}
\usepackage{amsmath}
\usepackage{amssymb}
\usepackage{graphicx}%
\newtheorem{theorem}{Theorem}

\newtheorem{corollary}[theorem]{Corollary}

\newtheorem{definition}[theorem]{Definition}
\newtheorem{example}[theorem]{Example}

\newtheorem{lemma}[theorem]{Lemma}

\newtheorem{proposition}[theorem]{Proposition}
\newtheorem{remark}[theorem]{Remark}

\newenvironment{proof}[1][Proof]{\noindent\textbf{#1.} }{\ \rule{0.5em}{0.5em}}
\begin{document}

\title{SPECTRAL\ AND\ REGULARITY\ \ PROPERTIES OF\ AN OPERATOR\ CALCULUS\ RELATED
TO\ LANDAU\ QUANTIZATION }
\author{Maurice de Gosson\thanks{M. de Gosson has been financed by the Austrian
Research Agency FWF (Projektnummer P20442-N13). } , Franz Luef\thanks{F. Luef
has been supported by the \textit{European Union EUCETIFA grant
MEXT-CT-2004-517154.}}\\\textit{Universit\"{a}t Wien}\\\textit{Fakult\"{a}t f\"{u}r Mathematik, NuHAG }\\\textit{Nordbergstrasse 15, AT-1090 Wien}}
\maketitle

\begin{abstract}
The theme of this work is that the theory of charged particles in a uniform
magnetic field can be generalized to a large class of operators if one uses an
extended a class of Weyl operators which we call "Landau--Weyl
pseudodifferential operators". The link between standard Weyl calculus and
Landau--Weyl calculus is made explicit by the use of an infinite family of
intertwining "windowed wavepacket transforms"; this makes possible the use of
the theory of modulation spaces to study various regularity properties. Our
techniques allow us not only to recover easily the eigenvalues and
eigenfunctions of the Hamiltonian operator of a charged particle in a uniform
magnetic field, but also to prove global hypoellipticity results and to study
the regularity of the solutions to Schr\"{o}dinger equations.

\end{abstract}
\tableofcontents

\section{Introduction}

The aim of this Communication is to compare the properties of a partial
differential operator (or, more generally, a Weyl pseudodifferential operator)
$A=a^{w}(x,-i\hbar\partial_{x})$ with those of the operator $\widetilde
{A}=a(X^{\gamma,\mu},Y^{\gamma,\mu})$ obtained by replacing formally $x$ and
$-i\hbar\partial_{x}$ by the vector fields
\begin{equation}
X^{\gamma,\mu}=\frac{\gamma}{2}x+\frac{i\hbar}{\mu}\partial_{y}\text{ \ ,
}Y^{\gamma,\mu}=\frac{\mu}{2}y-\frac{i\hbar}{\gamma}\partial_{x} \label{xjyj}%
\end{equation}
where $\gamma$ and $\mu$ are real scalars such that $\gamma\mu\neq0$. A
typical situation of physical interest\ is the following: choose for $A$ the
harmonic oscillator Hamiltonian
\begin{equation}
H_{\text{har}}=-\frac{\hbar^{2}}{2m}\frac{\partial^{2}}{\partial x^{2}}%
+\frac{m\omega^{2}}{2}x^{2} \label{har}%
\end{equation}
and $\gamma=1$, $\mu=m\omega$; defining the Larmor frequency $\omega
_{L}=\omega/2$ we obtain $H_{\text{har}}(X^{1,m\omega},Y^{1,m\omega
})=H_{\text{sym}}$ where
\begin{equation}
H_{\text{sym}}=-\frac{\hbar^{2}}{2m}\Delta_{x,y}-i\hbar\omega_{L}\left(
y\frac{\partial}{\partial x}-x\frac{\partial}{\partial y}\right)
+\frac{m\omega_{L}^{2}}{2}(x^{2}+y^{2}) \label{magn1}%
\end{equation}
is the Hamiltonian operator in the symmetric gauge of a charged particle
moving in the $x,y$ plane under the influence of constant magnetic field
orthogonal to the $x,y$ plane. Another interesting situation occurs if one
takes $\gamma=2$, $\mu=1$. In this case%
\[
H_{\text{har}}(X^{2,1},Y^{2,1})\Psi=H_{\text{har}}\star_{\hbar}\Psi
\]
where $\star_{\hbar}$ is the Moyal product familiar from deformation quantization.

Of course these observations are not of earthshaking importance unless we can
find a procedure for comparing the properties of both operators $A$ and
$\widetilde{A}$. In fact, as a rule, the initial operator $A$ is less
complicated that its counterpart $\widetilde{A}$ so one would like to deduce
the properties of the second from those of the first. For this we first have
to find a procedure allowing us to associate to a function $\psi\in
L^{2}(\mathbb{R}^{n})$ a function $\Psi\in L^{2}(\mathbb{R}^{2n});$ that
correspondence should be linear, and intertwine in some way the operators $A$
and $\widetilde{A}$; notice that the request for linearity excludes the choice
$\Psi=W\psi$ ($W$ the Wigner transform). It turns out that there exist many
procedures for transforming a function of, say, $x$ into a function of twice
as many variables; the Bargmann transform is an archetypical (and probably the
oldest) example of such a procedure. However, the Bargmann transform (and its
variants) is not sufficient to recover all the spectral properties of
$\widetilde{A}$ from those of $A$. For example, it is well known that the
\textquotedblleft Landau levels\textquotedblright\ of the magnetic operator
$H_{\text{sym}}$ are infinitely degenerate, so it is illusory to attempt to
recover the corresponding eigenvectors from those of $H_{\text{har}}$ (the
rescaled Hermite functions) using one single transform! This difficulty is of
course related to the fact that no isometry from $L^{2}(\mathbb{R}^{n})$ to
$L^{2}(\mathbb{R}^{2n})$ can take a basis of the first space to a basis of the
other, (intuitively $L^{2}(\mathbb{R}^{n})$ is \textquotedblleft much
smaller\textquotedblright\ than $L^{2}(\mathbb{R}^{2n})$). We will overcome
this difficulty by constructing an \textit{infinity} of isometries
$\mathcal{U}_{\phi}$ parametrized by the Schwartz space $\mathcal{S}%
(\mathbb{R}^{n})$; these isometries are defined in terms of the cross-Wigner
transform, or equivalently, in terms of the windowed Fourier transform
familiar from time-frequency and Gabor analysis. We will therefore call them
\textit{windowed wavepacket transforms}. For instance, in the case $\gamma=2$,
$\mu=1$ corresponding to the Moyal product, these isometries are (up to a
normalization factor) just the mappings $\psi\longmapsto\Psi=W(\psi,\phi)$
where $W(\psi,\phi)$ is the cross-Wigner transform.

One of the goals of this paper is to emphasize the great potential of a
particular class of function spaces, namely Feichtinger's modulation spaces,
whose elements can be defined in terms of decay properties of their windowed
Wigner transform; these spaces have turned out to be the proper setting for
the discussion of pseudodifferential operators in the last decade, and have
allowed to prove (or to recover) in an elementary way many results which would
otherwise requires the use of \textquotedblleft hard
analysis\textquotedblright. Although modulation spaces and their usefulness is
well-known in time-frequency an Gabor analysis they have not received a lot of
attention in quantum mechanics (they have however found some applications in
the study of Schr\"{o}dinger operators).

This article is structured as follows: in Sections \ref{sec1}--\ref{sec4} we
develop the theory in the case $\lambda=\mu=1$, that is we work with the
quantization rules%
\[
X=\frac{1}{2}x+i\hbar\partial_{y}\text{ \ , }Y=\frac{1}{2}y-i\hbar\partial_{x}%
\]
and their higher-dimensional generalizations%
\[
X_{j}=\frac{1}{2}x_{j}+i\hbar\partial_{y_{j}}\text{ \ , }Y_{j}=\frac{1}%
{2}y_{j}-i\hbar\partial_{x_{j}}.
\]
Furthermore we introduce the class of modulation spaces and recall some of
their basic properties that are of relevance in the present investigation. In
Section \ref{sec5} we show how the general case (\ref{xjyj}) and its
multi-dimensional generalization%
\[
X_{j}^{\gamma,\mu}=\frac{\gamma}{2}x_{{}}+\frac{i\hbar}{\mu}\partial_{y_{j}%
}\text{ \ , }Y_{j}^{\gamma,\mu}=\frac{\mu}{2}y_{j}-\frac{i\hbar}{\gamma
}\partial_{x_{j}}%
\]
can be reduced to this one. We thereafter apply the previous study to
deformation quantization.

\subsubsection*{Notation}

Functions (or distributions) on $\mathbb{R}^{n}$ will usually be denoted by
lower-case Greek letters $\psi,\phi...$ while functions (or distributions) on
$\mathbb{R}^{2n}$ will be denoted by upper-case Greek letters $\Psi,\Phi
,...$). We denote the inner product on $L^{2}(\mathbb{R}^{n})$ by $(\psi
|\phi)$ and the inner product on $L^{2}(\mathbb{R}^{2n})$ by $((\Psi|\Phi))$;
the associated norms are denoted by $||\psi||$ and $|||\Psi|||$, respectively.
Distributional brackets are denoted $\left\langle \cdot,\cdot\right\rangle $
in every dimension.

The standard symplectic form on the vector space $\mathbb{R}^{n}%
\times\mathbb{R}^{n}\equiv\mathbb{R}^{2n}$ is denoted by $\sigma$; it is
given, for $z=(x,y)$, $z^{\prime}=(x^{\prime},y^{\prime})$, by the formula
$\sigma(z,z^{\prime})=Jz\cdot z^{\prime}$ where $J=%
\begin{pmatrix}
0 & I\\
-I & 0
\end{pmatrix}
$. The symplectic group of $(\mathbb{R}^{2n},\sigma)$ is denoted by
$\operatorname*{Sp}(2n,\mathbb{R})$.

\section{Modulation Spaces: A Short Review\label{sec1}}

For a rather compete treatment of the theory of modulation spaces we refer to
Gr\"{o}chenig's book \cite{gr00}. The main point want to do in this section is
that, in contrast to the standard treatment of Weyl calculus, one can use with
profit Feichtinger's algebra $M^{1}(\mathbb{R}^{n})$ and its weighted variants
$M_{v_{s}}^{1}(\mathbb{R}^{n})$ as spaces of test functions instead of the
Schwartz class $\mathcal{S}(\mathbb{R}^{n})$. We also remark that a good class
of pseudodifferential symbols is provided by the modulation spaces $M_{v_{s}%
}^{1,\infty}(\mathbb{R}^{2n})$; it has recently been proved that they coincide
with the so-called \textit{Sj\"{o}strand classes}. In particular they contain
the H\"{o}rmander class $S_{0,0}^{0}(\mathbb{R}^{2n})$.

\subsection{Main definitions}

Roughly speaking, modulation spaces are characterized by the matrix
coefficients of the Schr\"{o}dinger representation of the Heisenberg group.
Recall that the \textit{Heisenberg--Weyl operators} $T(z_{0})$ are defined,
for $z_{0}=(x_{0},p_{0})\in\mathbb{R}^{2n}$ by:
\begin{equation}
T(z_{0})\psi(x)=e^{\frac{i}{\hbar}(y_{0}\cdot x-\frac{1}{2}y_{0}\cdot x_{0}%
)}\psi(x-x_{0}). \label{HW}%
\end{equation}
We have
\begin{equation}
T(z_{1}+z_{2})=e^{-\frac{i}{2\hbar}\sigma(z_{1},z_{2})}T(z_{1})T(z_{2}%
)~~z_{1},z_{2}\in\mathbb{R}^{2n} \label{tzz}%
\end{equation}
hence $z\longmapsto T(z)$ is a projective representation of $\mathbb{R}^{2n}$,
it is called the Schr\"{o}dinger representation of the Heisenberg group. Let
$\psi,g$ be in $L^{2}(\mathbb{R}^{n})$. Then the matrix coefficient of the
Schr\"{o}dinger representation is given by
\begin{equation}
V_{\phi}^{\hbar}\psi(z_{0})=\langle\psi,T(z_{0})\phi\rangle=e^{\frac{i}%
{2\hbar}y_{0}\cdot x_{0}}\int_{\mathbb{R}^{n}}e^{\frac{i}{\hbar}y_{0}\cdot
x}\psi(x)\overline{\phi(x-x_{0})}dx. \label{vh}%
\end{equation}
When $\hbar=1/2\pi$ it is, up to the exponential in front of the integral, the
\textit{short-time Fourier transform} (STFT) $V_{\phi}\psi$ familiar from
time-frequency analysis:%
\begin{equation}
V_{\phi}\psi(z_{0})=\int_{\mathbb{R}^{n}}e^{2\pi iy_{0}\cdot x}\psi
(x)\overline{\phi(x-x_{0})}dx \label{STFT}%
\end{equation}
($V_{\phi}\psi$ is also called the \textquotedblleft voice
transform\textquotedblright\ or \textquotedblleft sliding
transform\textquotedblright\ in signal theory).

In what follows $p$ is a non-negative real number and $m$ a real weight
function on $\mathbb{R}^{2n}$. By definition a distribution $\psi
\in\mathcal{S}^{\prime}(\mathbb{R}^{n})$ is in the modulation space $M_{m}%
^{p}(\mathbb{R}^{n})$ if there exists $\phi\in\mathcal{S}(\mathbb{R}^{n})$,
$\phi\neq0$, such that $V_{\phi}\psi\in L_{m}^{p}(\mathbb{R}^{2n})$, that is
if
\begin{equation}
||\psi||_{M_{m}^{p}}=\left(  \int|V_{\phi}\psi(z)|^{p}m(z)dz\right)
^{1/p}<\infty. \label{norm}%
\end{equation}
The essential point is that that the definition above is independent of the
choice of the window $\phi$: if it holds for one such window, it holds for
all. Formula (\ref{norm}) defines a norm on $M_{m}^{p}(\mathbb{R}^{n})$ and
the replacement of $\phi$ by another window leads to an equivalent norm.
Modulation spaces are Banach spaces for the topology defined by (\ref{norm}).
(See \cite{fe83-4,gr00} for the general theory).

\subsection{Two examples}

In what follows $\left\langle \cdot\right\rangle ^{s}$ denotes the weight
function defined by
\begin{equation}
\left\langle z\right\rangle ^{s}=(1+|z|^{2})^{s/2}\text{.} \label{wz}%
\end{equation}

\begin{itemize}
\item \textbf{The modulation spaces} $M_{v_{s}}^{1}(\mathbb{R}^{n})$.
\end{itemize}

They are defined as follows: for $s\geq0$ we consider the weighted
$L^{1}(\mathbb{R}^{2n})$ space
\begin{equation}
L_{v_{s}}^{1}(\mathbb{R}^{2n})=\{\Psi\in\mathcal{S}^{\prime}(\mathbb{R}%
^{2n}):||\Psi||_{L_{v_{s}}^{1}}<\infty\} \label{defl1}%
\end{equation}
where $||\cdot||_{L_{v_{s}}^{1}}$ is the norm given by
\begin{equation}
||\Psi||_{L_{v_{s}}^{1}}^{2}=\int_{\mathbb{R}^{2n}}|\Psi(z)|\left\langle
z\right\rangle ^{s}dz. \label{norm1}%
\end{equation}
The corresponding modulation space is%
\begin{equation}
M_{v_{s}}^{1}(\mathbb{R}^{n})=\{\psi\in\mathcal{S}^{\prime}(\mathbb{R}%
^{n}):V_{\phi}\psi\in L_{v_{s}}^{1}(\mathbb{R}^{2n})\} \label{m1s}%
\end{equation}
where $V_{\phi}$ is the STFT (\ref{STFT}) and $\phi$ is an \emph{arbitrary}
element of the Schwartz space $\mathcal{S}(\mathbb{R}^{n})$; the formula%
\begin{equation}
||\psi||_{M_{v_{s}}^{1}}^{\phi}=||V_{\phi}\psi||_{L_{v_{s}}^{1}}
\label{norm1s}%
\end{equation}
defines a norm on $M_{v_{s}}^{1}(\mathbb{R}^{n})$ and if we change $\phi$ into
another element of $\mathcal{S}(\mathbb{R}^{n})$ we obtain an equivalent norm
(Proposition 12.1.2 (p.246) in \cite{gr00}). The space $M_{v_{s}}%
^{1}(\mathbb{R}^{n})$ can also be very simply described in terms of the Wigner
transform: we have
\begin{equation}
M_{v_{s}}^{1}(\mathbb{R}^{n})=\{\psi\in\mathcal{S}^{\prime}(\mathbb{R}%
^{n}):W\psi\in L_{v_{s}}^{1}(\mathbb{R}^{2n})\}. \label{mis}%
\end{equation}
In the case $s=0$ we obtain the \textit{Feichtinger algebra }$M^{1}%
(\mathbb{R}^{n})=M_{0}^{1}(\mathbb{R}^{n})$. We have the inclusions%
\begin{equation}
\mathcal{S}(\mathbb{R}^{n})\subset M^{1}(\mathbb{R}^{n})\subset C^{0}%
(\mathbb{R}^{n})\cap L^{1}(\mathbb{R}^{n})\cap L^{2}(\mathbb{R}^{n}).
\label{incl}%
\end{equation}
The Feichtinger algebra can be used with profit as a space of test functions;
its dual Banach space
\begin{equation}
M^{\infty}(\mathbb{R}^{n})=\{\psi\in\mathcal{S}^{\prime}(\mathbb{R}^{n}%
):\sup_{z\in\mathbb{R}^{2n}}\left(  \left\langle z\right\rangle ^{N}|V_{\phi
}\psi|\right)  <\infty\text{ for all }N\} \label{minf}%
\end{equation}
contains the Dirac distribution $\delta$. Furthermore $M^{1}(\mathbb{R}^{n})$
is the smallest Banach space invariant under Heisenberg--Weyl operators (it is
the space of integrable vectors of the Heisenberg--Weyl representation). Note
that step functions are not in $M^{1}(\mathbb{R}^{n})$ but triangle functions
(which are the convolutions of two step functions) are. For these reasons the
modulation spaces $M_{v_{s}}^{1}(\mathbb{R}^{n})$ are considerably larger
classes of test functions than the Schwartz space $\mathcal{S}(\mathbb{R}%
^{n})$.

The spaces $M_{v_{s}}^{1}(\mathbb{R}^{n})$, besides other properties, are
invariant under Fourier transform and more generally, under the action of the
metaplectic group. They are also preserved by rescalings:

\begin{lemma}
\label{lemsca}Let $\lambda$ be a real number different from zero and set
$\psi_{\lambda}(x)=\psi(\lambda x)$. We have $\psi_{\lambda}\in M_{v_{s}}%
^{p}(\mathbb{R}^{n})$ if and only if $\psi\in M_{v_{s}}^{p}(\mathbb{R}^{n})$.
\end{lemma}

\begin{proof}
It immediately follows from definition (\ref{STFT}) of the STFT\ that we have%
\begin{equation}
V_{\phi}\psi_{\lambda}(z)=\lambda^{-n}V_{\phi_{1/\lambda}}\psi(z/\lambda)
\label{vlambda}%
\end{equation}
hence, performing a simple change of variable,
\[
\int_{\mathbb{R}^{2n}}|V_{\phi}\psi_{\lambda}(z)|^{p}\left\langle
z\right\rangle ^{s}dz=\lambda^{n}\int_{\mathbb{R}^{2n}}|V_{\phi_{1/\lambda}%
}\psi(z)|^{p}\left\langle \lambda z\right\rangle ^{s}dz.
\]
Since $\left\langle \lambda z\right\rangle ^{s}\leq(1+\lambda^{2}%
)^{s/2}\left\langle z\right\rangle ^{s}$ it follows that there exists a
constant $C_{\lambda}>0$ such that%
\[
\int_{\mathbb{R}^{2n}}|V_{\phi}\psi_{\lambda}(z)|^{p}\left\langle
z\right\rangle ^{s}dz\leq C_{\lambda}\int_{\mathbb{R}^{2n}}|V_{\phi
_{1/\lambda}}\psi(z)|^{p}\left\langle z\right\rangle ^{s}dz;
\]
the integral in the right hand side is convergent if and only if $\psi\in
M_{v_{s}}^{p}(\mathbb{R}^{n})$, proving the necessity of the condition. That
the condition is sufficient follows replacing $\psi_{\lambda}$ by $\psi$ in
the argument above.
\end{proof}

\begin{itemize}
\item \textbf{The modulation spaces} $M_{v_{s}}^{1,\infty}(\mathbb{R}^{2n})$.
\end{itemize}

They are defined by the condition $a\in M_{v_{s}}^{1,\infty}(\mathbb{R}^{2n})$
if and only if we have%
\[
||a||_{M_{v_{s}}^{1,\infty}}^{\Phi}=\int_{\mathbb{R}^{2n}}\sup_{z\in
\mathbb{R}^{2n}}|V_{\Phi}a(z,\zeta)|\left\langle \zeta\right\rangle
^{-s}d\zeta<\infty
\]
for one (and hence all) $\Phi\in\mathcal{S}(\mathbb{R}^{2n})$; here $V_{\Phi}$
is the STFT of functions on $\mathbb{R}^{2n}$. The use of the letter $a$ for
the elements of $M_{v_{s}}^{1,\infty}(\mathbb{R}^{2n})$ suggests that this
space could be used as a symbol class. This is indeed the case; we mention
that when $s=0$ one recovers the so-called \textquotedblleft Sj\"{o}strand
classes\textquotedblright\ \cite{sj94,sj95} which have been studied and
developed by Gr\"{o}chenig \cite{gr06,gr06bis} from the point of view of
modulation space theory. As a symbol class is rather large; for instance%
\begin{equation}
C^{2n+1}(\mathbb{R}^{2n})\subset M^{1,\infty}(\mathbb{R}^{2n})=M_{0}%
^{1,\infty}(\mathbb{R}^{2n}). \label{c2n}%
\end{equation}

The following result, follows from Theorem 4.1 and its Corollary 4.2 in
\cite{gr06bis}; it clearly demonstrates the usefulness of the spaces
$M_{v_{s}}^{1,\infty}(\mathbb{R}^{2n})$ in the theory of pseudodifferential operators:

\begin{proposition}
\label{propm1s}(i) An operator $A$ with Weyl symbol $a\in M_{v_{s}}^{1,\infty
}(\mathbb{R}^{2n})$ is bounded on every modulation space $M_{v_{s}}%
^{1}(\mathbb{R}^{n})$; (ii) If $a\in M_{0}^{1,\infty}(\mathbb{R}^{2n})$ then
$A$ maps $L^{1}(\mathbb{R}^{n})$ into Feichtinger's algebra $M_{0}%
^{1}(\mathbb{R}^{n})$.
\end{proposition}

\noindent(It is actually proven in \cite{gr06bis} that $A$ is bounded on any
modulation space $M_{m}^{p,q}(\mathbb{R}^{n})$ when $m$ is an arbitrary
moderate weight).

\subsection{Duality and kernel theorems\label{subseckern}}

The dual of $M_{v_{s}}^{1}(\mathbb{R}^{n})$ is the Banach space $M_{1/v_{s}%
}^{\infty}(\mathbb{R}^{n})$ with the norm
\begin{equation}
\Vert\psi\Vert_{M_{1/v_{s}}^{\infty}}=\mathrm{sup}_{z\in\mathbb{R}^{2n}%
}\left(  V_{\phi}\psi(z)|\left\langle z\right\rangle ^{s}\right)  <\infty.
\end{equation}
Note that the Schwartz space $\mathcal{S}(\mathbb{R}^{n})$ is the projective
limit of the spaces $\{M_{v_{s}}^{1}(\mathbb{R}^{n}):s\geq0\}$ and
consequently $\mathcal{S}^{\prime}(\mathbb{R}^{n})$ has a description as
inductive limit of the spaces $\{M_{1/v_{s}}^{\infty}(\mathbb{R}^{n}%
):s\geq0\}$, i.e.
\begin{equation}
\mathcal{S}(\mathbb{R}^{n})=\bigcap_{s\geq0}M_{v_{s}}^{1}(\mathbb{R}%
^{n})~~\text{and}~~\mathcal{S}^{\prime}(\mathbb{R}^{n})=\bigcup_{s\geq
0}M_{1/v_{s}}^{\infty}(\mathbb{R}^{n}) \label{30}%
\end{equation}
with $(\Vert.\Vert_{M_{v_{s}}^{1}})_{s\geq0}$ and $(\Vert.\Vert_{M_{1/v_{s}%
}^{\infty}})_{s\geq0}$ as family of seminorms for $\mathcal{S}(\mathbb{R}%
^{n})$ and $\mathcal{S}^{\prime}(\mathbb{R}^{n})$, respectively.

Therefore results about the class of modulation spaces $M_{v_{s}}%
^{1}(\mathbb{R}^{n})$ and its dual $M_{1/v_{s}}^{\infty}(\mathbb{R}^{n})$
translate into corresponding results about the Schwartz class $\mathcal{S}%
(\mathbb{R}^{n})$ and the tempered distributions $\mathcal{S}^{\prime
}(\mathbb{R}^{n})$. The great relevance of the pair $\mathcal{S}%
(\mathbb{R}^{n})$ and $\mathcal{S}^{\prime}(\mathbb{R}^{n})$ comes from the
kernel theorem and Feichtinger showed that for the pair of modulation spaces
$M_{v_{s}}^{1}(\mathbb{R}^{n})$ and $M_{1/v_{s}}^{\infty}(\mathbb{R}^{n})$
there also exists a kernel theorem (see \cite{gr00}, \S 14.4, for a short proof):

\begin{theorem}
[Kernel theorem]\label{KernelTheorem} Let $A$ be a continuous operator
$M_{{v_{s}}}^{1}(\mathbb{R}^{n})\longrightarrow M_{1/v_{s}}^{\infty
}(\mathbb{R}^{n})$. There exists a distribution $\mathcal{K}_{A}\in
M_{1/v_{s}}^{\infty}(\mathbb{R}^{2n})$ such that
\begin{equation}
\langle A\psi,\phi\rangle=\langle\mathcal{K}_{A},\psi\otimes\overline{\phi
}\rangle~~\text{for}~~\phi,\psi\in M_{v_{s}}^{1}(\mathbb{R}^{n}).
\end{equation}

\end{theorem}

As a consequence, using the intersections (\ref{30}) we get the following
version of the Schwartz kernel theorem.

\begin{theorem}
Let $A$ be a continuous linear operator from $\mathcal{S}(\mathbb{R}^{n})$ to
$\mathcal{S}^{\prime}(\mathcal{R}^{n})$. Then $A$ extends to a bounded
operator from $M_{v_{s}}^{1}(\mathbb{R}^{n})$ to $M_{1/v_{s}}^{\infty
}(\mathbb{R}^{n})$ for some $s\geq0$.
\end{theorem}

Therefore our framework covers the traditional setting of pseudo-differential calculus.

\section{Landau--Weyl Calculus}

For the reader's convenience we begin by quickly reviewing the basics of
standard Weyl calculus. See for instance \cite{hor1,sh87} for details and proofs.

\subsection{Review of standard Weyl calculus}

In view of the kernel theorem above, there exists, for every linear continuous
operator $A:M_{v_{s}}^{1}(\mathbb{R}^{n})\longrightarrow M_{1/v_{s}}^{\infty
}(\mathbb{R}^{n})$, a distribution $\mathcal{K}_{A}\in M_{1/v_{s}}^{\infty
}(\mathbb{R}^{n}\times\mathbb{R}^{n})$ such that $A\psi(x)=\left\langle
\mathcal{K}_{A}(x,\cdot),\psi\right\rangle $ for every $\psi\in M_{v_{s}}%
^{1}(\mathbb{R}^{n})$.

By definition the contravariant symbol of $A$ is the distribution $a$ defined
by the Fourier transform
\begin{equation}
a(x,y)=\int_{\mathbb{R}^{n}}e^{-\frac{i}{\hbar}y\cdot\eta}\mathcal{K}%
_{A}(x+\tfrac{1}{2}\eta,x-\tfrac{1}{2}\eta)d\eta\label{axy}%
\end{equation}
(the integral is interpreted in the distributional sense) and we can thus
write formally
\begin{equation}
A\psi(x)=\left(  \tfrac{1}{2\pi\hbar}\right)  ^{n/2}\iint\nolimits_{\mathbb{R}%
^{n}\times\mathbb{R}^{n}}e^{\frac{i}{\hbar}y(x-x^{\prime})}a(\tfrac{1}%
{2}(x+x^{\prime}),y)\psi(x^{\prime})dx^{\prime}dy; \label{apsosc}%
\end{equation}
strictly speaking this formula only makes sense when $a\in M_{v_{s}}%
^{1}(\mathbb{R}^{2n})$; when this is not the case the double integral has to
be reinterpreted in some way, for instance as a repeated or \textquotedblleft
oscillatory\textquotedblright\ integral; see for instance Tr\`{e}ves' book
\cite{Treves} for an exposition of various techniques which are useful in this context.

By definition the covariant symbol $a_{\sigma}$ of $A$ is the symplectic
Fourier transform
\begin{equation}
a_{\sigma}(z)=F_{\sigma}a(z)=\left(  \tfrac{1}{2\pi\hbar}\right)
^{n}\left\langle e^{-\frac{i}{\hbar}\sigma(z,\cdot)},a(\cdot)\right\rangle .
\label{sft}%
\end{equation}
Using the covariant symbol we can rewrite (\ref{apsosc}) as an operator-valued
(Bochner) integral%
\begin{equation}
A=\left(  \tfrac{1}{2\pi\hbar}\right)  ^{n}\int_{\mathbb{R}^{2n}}a_{\sigma
}(z)T(z)dz \label{bochner}%
\end{equation}
where $T(z)$ is the Heisenberg--Weyl operator (\ref{HW}).

Note that Weyl operators are composed in the following way. Assume that $A$
and $B$ are mappings on $M_{v_{s}}^{1}(\mathbb{R}^{n})$; then $C=BA$ is
defined and its contravariant and covariant symbols are given by the formulae
\begin{align}
c(z)  &  =\left(  \tfrac{1}{4\pi\hbar}\right)  ^{2n}\iint\nolimits_{\mathbb{R}%
^{2n}\times\mathbb{R}^{2n}}e^{\frac{i}{2\hbar}\sigma(u,v)}a(z+\tfrac{1}%
{2}u)b(z-\tfrac{1}{2}v)dudv\label{compo1}\\
c_{\sigma}(z)  &  =\left(  \tfrac{1}{2\pi\hbar}\right)  ^{n}\int
_{\mathbb{R}^{2n}}e^{\frac{i}{2\hbar}\sigma(z,z^{\prime})}a_{\sigma
}(z-z^{\prime})b_{\sigma}(z^{\prime})dz^{\prime}. \label{compo2}%
\end{align}

The last two equations have a natural interpretation in terms of involutive
representations of the twisted group algebra $L^{1}(\mathbb{R}^{2n},\chi)$ for
the $2$-cocycle $\chi(z,z^{\prime})=e^{\frac{i}{2\hbar}\sigma(z,z^{\prime})}$.
Namely, the unitary representation of the Heisenberg group by the
Heisenberg--Weyl operators $T(z)$ gives an involutive faithful representation
of $L^{1}(\mathbb{R}^{2n},\chi)$ via
\begin{equation}
\pi_{\text{int}}(a)=\int_{\mathbb{R}^{2n}}a(z)T(z)dz
\end{equation}
for $a\in L^{1}(\mathbb{R}^{2n})$. In representation theory $\pi_{\text{int}}$
is called the \textit{integrated representation} of the representation $T(z)$.
The product of $\pi_{\text{int}}(a)$ and $\pi_{\text{int}}(b)$ for $a,b\in
L^{1}(\mathbb{R}^{2n})$ yields another element $\pi_{\text{int}}(c)$ of
$L^{1}(\mathbb{R}^{2n},\chi)$, where $c$ is obtained from $a$ and $b$ by
\textit{twisted convolution}:
\begin{equation}
c(z)=a\natural b(z)=\left(  \tfrac{1}{2\pi\hbar}\right)  ^{n}\int
_{\mathbb{R}^{2n}}e^{\frac{i}{2\hbar}\sigma(z,z^{\prime})}a_{\sigma
}(z-z^{\prime})b_{\sigma}(z^{\prime})dz^{\prime}%
\end{equation}
(see for instance \cite{Wong}). Consequently, the composition of two operators
in the Weyl calculus is actually the twisted convolution of their covariant
symbols in the twisted group algebra $L^{1}(\mathbb{R}^{2n},\chi)$.

Two particularly nice features of the Weyl pseudodifferential calculus are the following:

\begin{itemize}
\item Assume that $A:M_{v_{s}}^{1}(\mathbb{R}^{n})\longrightarrow M_{1/v_{s}%
}^{\infty}(\mathbb{R}^{n})$. Then the contravariant symbol of $A^{\ast}$ is
complex conjugate to that of $A$. In particular, the Weyl operator $A$ is
self-adjoint if and only if its contravariant symbol $a$ is real;

\item For every $s\in\operatorname*{Sp}(2n,\mathbb{R})$ there exists a unitary
operator $S$, uniquely defined up to a complex factor, such that the Weyl
operator $B$ with contravariant symbol $b=a\circ s^{-1}$ is given by the
formula
\begin{equation}
B=SAS^{-1}; \label{metaco}%
\end{equation}
$S$ can be chosen as a multiple of any of the two metaplectic operators
covering $s$.
\end{itemize}

We recall that the metaplectic group $\operatorname*{Mp}(2n,\mathbb{R})$ is a
faithful unitary representation of the twofold connected covering of
$\operatorname*{Sp}(2n,\mathbb{R})$; for a detailed account of the well-known
(and less well-known) properties of $\operatorname*{Mp}(2n,\mathbb{R})$ see
\cite{Birk}, Chapter 7.

\subsection{Landau--Weyl operators}

In our discussion of the Weyl calculus we stressed its interpretation in terms
of the integrated representation of the Heisenberg--Weyl operators. A basic
result about the representation theory of groups says that there is a
one-to-one correspondence between projective representations of a group and
integrated representations of the twisted group algebra of the group.
Therefore a new representation of the Heisenberg group yields a new kind of
calculus for pseudo-differential operators. In a series of papers on phase
space Schr\"{o}dinger equations the first author has implicitly made use of
this fact. In the following we want to present these results in terms of
integrated representation of a representation of the Heisenberg group on
$L^{2}(\mathbb{R}^{2n})$.

We define unitary operators $\widetilde{T}(z)$ on $L^{2}(\mathbb{R}^{2n})$ by
the formula
\begin{equation}
\widetilde{T}(z_{0})\Psi(z)=e^{-\tfrac{i}{2\hbar}\sigma(z,z_{0})}\Psi
(z-z_{0}). \label{thw}%
\end{equation}
We point out that these operators satisfy the relation%
\[
\widetilde{T}(z_{0}+z_{1})=e^{-\tfrac{i}{2\hbar}\sigma(z,z_{0})}\widetilde
{T}(z_{0})\widetilde{T}(z_{1})
\]
which is formally similar to the relation (\ref{tzz}) for the Heisenberg--Weyl
operators (\ref{HW}). In fact, in \cite{Birk}, Chapter 10, one of us has shown
that these operators can be used to construct an unitary irreducible
representation of the Heisenberg group $H_{n}$ on (infinitely many) closed
subspace(s) of $L^{2}(\mathbb{R}^{2n})$. This is achieved using the wavepacket
transforms we define in Section \ref{sec2}.

We now \emph{define} the operator $\widetilde{A}:\mathcal{S}(\mathbb{R}%
^{2n})\longrightarrow\mathcal{S}^{\prime}(\mathbb{R}^{2n})$ by replacing
$T(z)$ with $\widetilde{T}(z)$ in formula (\ref{bochner}):
\begin{equation}
\widetilde{A}=\left(  \tfrac{1}{2\pi\hbar}\right)  ^{n}\int_{\mathbb{R}^{2n}%
}{a}_{\sigma}(z)\widetilde{T}(z)dz, \label{ahwbis}%
\end{equation}
i.e. as integrated representation of $\widetilde{T}(z)$.

\begin{definition}
We will call the operator $\widetilde{A}$ defined by (\ref{ahwbis}) the
\textit{Landau--Weyl (for short: LW) operator} with symbol ${a}$ (or:
associated with the Weyl operator $A$).
\end{definition}

This terminology is motivated by the fact that the magnetic operator
(\ref{magn1}) appears as a particular case of these operators, choosing for
$a$ the harmonic oscillator Hamiltonian. Let us in fact determine the
contravariant symbol of $\widetilde{A}$, viewed as Weyl operator $M_{v_{s}%
}^{1}(\mathbb{R}^{2n})\longrightarrow M_{1/v_{s}}^{\infty}(\mathbb{R}^{2n})$.
Because of the importance of this result for the rest of this paper we give it
the status of a theorem:

\begin{theorem}
\label{proplww}Let $a$ be the contravariant symbol of the Weyl operator $A$.
Let $(z,\zeta)\in\mathbb{R}^{2n}\times\mathbb{R}^{2n}$ and $z=(x,y)$,
$\zeta=(p_{x},p_{y})$. The contravariant symbol of $\widetilde{A}$, viewed as
a Weyl operator $M_{v_{s}}^{1}(\mathbb{R}^{2n})\longrightarrow M_{1/v_{s}%
}^{\infty}(\mathbb{R}^{2n})$ is given by the formula%
\begin{equation}
\widetilde{a}(z,\zeta)=a(\tfrac{1}{2}z-J\zeta)=a(\tfrac{1}{2}x-p_{y},\tfrac
{1}{2}y+p_{x}). \label{azzu}%
\end{equation}

\end{theorem}

\begin{proof}
The kernel of $\widetilde{A}$ is given by the formula
\begin{equation}
\mathcal{K}_{\widetilde{A}}(z,u)=\left(  \tfrac{1}{2\pi\hbar}\right)
^{n/2}e^{\frac{i}{2\hbar}\sigma(z,u)}a_{\sigma}(z-u) \label{katilde}%
\end{equation}
as is easily seen by performing the change of variables $u=z-z_{0}$ in
definition (\ref{ahwbis}) and noting that $\sigma(z,z-u)=-\sigma(z,u)$. We
have (cf. formula (\ref{axy}))%
\[
\widetilde{a}(z,\zeta)=\int_{\mathbb{R}^{2n}}e^{-\frac{i}{\hbar}\zeta\cdot
\eta}\mathcal{K}_{\widetilde{A}}(z+\tfrac{1}{2}\eta,z-\tfrac{1}{2}\eta)d\eta
\]
hence, using the identity $\sigma(z+\tfrac{1}{2}\eta,z-\tfrac{1}{2}%
\eta)=-\sigma(z,\eta)$,
\begin{equation}
\widetilde{a}(z,\zeta)=\left(  \tfrac{1}{2\pi\hbar}\right)  ^{n}%
\int_{\mathbb{R}^{2n}}e^{-\frac{i}{\hbar}\zeta\cdot\eta}e^{-\frac{i}{2\hbar
}\sigma(z,\eta)}a_{\sigma}(\eta)d\eta\label{az}%
\end{equation}
where $a_{\sigma}$ is the covariant symbol of $A$. By definition (\ref{sft})
of the symplectic Fourier transform we have
\[
e^{-\frac{i}{\hbar}\zeta\cdot\eta}a_{\sigma}(\eta)=\left(  \tfrac{1}{2\pi
\hbar}\right)  ^{n}\int_{\mathbb{R}^{2n}}e^{-\frac{i}{\hbar}\zeta\cdot\eta
}e^{-\frac{i}{\hbar}\sigma(\eta,z)}a(z)dz
\]
hence, observing that $\sigma(\eta,z)+\zeta\cdot\eta=\sigma(\eta,z+J\zeta)$,
\[
e^{-\frac{i}{\hbar}\zeta\cdot\eta}a_{\sigma}(\eta)=\left(  \tfrac{1}{2\pi
\hbar}\right)  ^{n}\int_{\mathbb{R}^{2n}}e^{-\frac{i}{\hbar}\sigma(\eta
,z)}T_{J\zeta}a(z)dz
\]
where $T_{J\zeta}a(z)=a(z-J\zeta)$, that is%
\[
e^{-\frac{i}{\hbar}\zeta\cdot\eta}a_{\sigma}(\eta)=F_{\sigma}(T_{J\zeta
}a)(\eta).
\]
Formula (\ref{az}) can thus be rewritten as%
\[
\widetilde{a}(2z,\zeta)=\left(  \tfrac{1}{2\pi\hbar}\right)  ^{n}%
\int_{\mathbb{R}^{2n}}e^{-\frac{i}{\hbar}\sigma(z,\eta)}F_{\sigma}(T_{J\zeta
}a)(\eta)d\eta
\]
hence $\widetilde{a}(2z,\zeta)=T_{J\zeta}a(z)$ (the symplectic Fourier
transform is involutive); formula (\ref{azzu}) follows.
\end{proof}

Here are two immediate consequences of the result above. The first says that
$\widetilde{A}$ is self-adjoint if and only $A$ is; the second says that the
LW operators compose as the usual Weyl operators.

\begin{corollary}
\label{corab}(i) The operator $\widetilde{A}$ is self-adjoint if and only $a$
is real; (ii) The contravariant symbol of $\widetilde{C}=\widetilde
{A}\widetilde{B}$ is given by $\widetilde{c}(z)=c(\tfrac{1}{2}z-J\zeta)$ where
$c$ is the contravariant symbol of $C=AB$.
\end{corollary}

\begin{proof}
To prove property (i) it suffices to note that $\widetilde{a}$ is real if and
only $a$ is. Property (ii) immediately follows from (\ref{compo1}) and
(\ref{azzu}).
\end{proof}

Another consequence of these results is a statement about Landau--Weyl
operators on $M_{v_{s}}^{1}(\mathbb{R}^{n})$. A well-known result due to
Feichtinger (see Gr\"{o}chenig \cite{gr96} for a proof) asserts that a Weyl
pseudodifferential operator bounded on $M_{v_{s}}^{1}(\mathbb{R}^{n})$ is of
trace-class. Consequently the same results holds for operators in the
Landau--Weyl calculus. Therefore we can compute the trace of these operators
by integrating their kernel along the diagonal

\begin{corollary}
Let $\widetilde{A}$ be a bounded selfadjoint operator on $M_{v_{s}}%
^{1}(\mathbb{R}^{n})$ with kernel $\mathcal{K}_{\widetilde{A}}$. The trace of
$\widetilde{A}$ is given by
\begin{equation}
\mathrm{Tr}(\widetilde{A})=\int_{\mathbb{R}^{2n}}\mathcal{K}_{\widetilde{A}%
}(z,z)dz. \label{trace}%
\end{equation}

\end{corollary}

\begin{remark}
We emphasize that trace formulas of the type (\ref{trace}) are usually not
true for arbitrary trace-class operators (see the very relevant discussion of
\textquotedblleft trace formulas\textquotedblright\ in Reed and Simon
\cite{RS}.)
\end{remark}

\subsection{Symplectic covariance and metaplectic operators}

The symplectic covariance property (\ref{metaco}) carries over to the LW
calculus: for every $\widetilde{S}\in\operatorname*{Mp}(4n,\mathbb{R})$ with
projection $s$ we have
\begin{equation}
\widetilde{S}\widetilde{T}(z)\widetilde{S}^{-1}=\widetilde{T}(sz)\text{ \ ,
\ }\widetilde{S}\widetilde{A}\widetilde{S}^{-1}=\widetilde{B} \label{sas}%
\end{equation}
where $\widetilde{B}$ corresponds to $b=a\circ s^{-1}$.

Metaplectic operators are Weyl operators in their own right (see
\cite{lettmp,Birk}). Let us determine the corresponding LW operators.

\begin{proposition}
Let $S\in\operatorname*{Mp}(2n,\mathbb{R})$ have projection $s\in
\operatorname*{Sp}(2n,\mathbb{R})$. If $\det(s-I)\neq0$ then
\begin{equation}
\widetilde{S}=\left(  \tfrac{1}{2\pi\hbar}\right)  ^{n}\int_{\mathbb{R}^{2n}%
}a_{\sigma}^{S}(z)\widetilde{T}(z)dz \label{mplw}%
\end{equation}
where the function $a_{\sigma}^{S}$ is given by%
\begin{equation}
a_{\sigma}^{S}(z)=\frac{i^{\nu(S)}}{\sqrt{|\det(s-I)|}}\exp\left(  \frac
{i}{2\hbar}M_{s}z\cdot z\right)  \label{mpweyl}%
\end{equation}
with $M_{s}=\tfrac{1}{2}J(s+I)(s-I)^{-1}M_{s}^{T}$. The integer $\nu(S)$ is
the class modulo $4$ of the Conley--Zehnder index \cite{JMP,RMP}of a path
joining the identity to $S$ in $\operatorname*{Sp}(2n,\mathbb{R})$.
\end{proposition}

\begin{proof}
In \cite{lettmp} one of us showed that every $S\in\operatorname*{Mp}%
(2n,\mathbb{R})$ with $\det(s-I)\neq0$ can be written in the form \
\[
S=\left(  \tfrac{1}{2\pi\hbar}\right)  ^{n}\int_{\mathbb{R}^{2n}}a_{\sigma
}^{S}(z)T(z)dz
\]
where $a_{\sigma}^{S}$ is given by (\ref{mpweyl}). Property (i) follows.
\end{proof}

The operators $\widetilde{S}$ are metaplectic operators belonging to
$\operatorname*{Mp}(4n,\mathbb{R})$; we will not prove this fact here, but
rather focus on a class of elementary operators which will be very useful for
defining the general parameter dependent LW calculus in Section \ref{sec5}:

\begin{lemma}
\label{lemu}For $(\gamma,\mu)\in\mathbb{R}^{2}$, $\gamma\mu\neq0$, let
$\widetilde{S}^{\gamma,\mu}$ be the unitary operator on $L^{2}(\mathbb{R}%
^{2n})$ defined by
\begin{equation}
\widetilde{S}^{\gamma,\mu}\Psi(x,y)=|\gamma\mu|^{n}\Psi(\gamma x,\mu y).
\label{slu}%
\end{equation}
We have $\widetilde{S}^{\gamma,\mu}\in\operatorname*{Mp}(4n,\mathbb{R})$,
$(\widetilde{S}^{\gamma,\mu})^{-1}=\widetilde{S}^{\frac{1}{\gamma},\frac
{1}{\mu}}$, and the projection of $\widetilde{S}^{\gamma,\mu}$ onto
$\operatorname*{Sp}(4n,\mathbb{R})$ is the diagonal matrix
\begin{equation}
S^{\gamma,\mu}=\operatorname*{diag}(\gamma^{-1}I,\mu^{-1}I,\gamma I,\mu I)
\label{sdiag}%
\end{equation}
($I$ the n$\times n$ identity).
\end{lemma}

\begin{proof}
That $\widetilde{S}^{\gamma,\mu}\in\operatorname*{Mp}(4n,\mathbb{R})$ and
formula (\ref{sdiag}) are standard results from the theory of metaplectic
operators \cite{Birk,GS1,GS2}.
\end{proof}

\section{Windowed Wavepacket Transforms\label{sec2}}

In this section we treat \textit{resolutions of identity} from a
representation theoretic point of view. This approach has been of great
relevance in various works in mathematics and physics (see e.g.
\cite{fegr89,si80} for a very general discussion of the topic). In the
terminology of \cite{si80} we investigate in the present section coherent
vectors and coherent projections generated by the square integrable
representations $T(z)$ and $\tilde{T}(z)$. The square-integrability of the
representation $T(z)$ of the Heisenberg group on $L^{2}(\mathbb{R}^{n})$ is
the \textit{Moyal identity:}
\begin{equation}
((V_{\phi_{1}}\psi_{1}|V_{\phi_{2}}\psi_{2}))=(\psi_{1}|\psi_{2}%
)\overline{(\phi_{1}|\phi_{2})} \label{Moyal}%
\end{equation}
for $\phi_{1},\phi_{2},\psi_{1},\psi_{2}$ in $L^{2}(\mathbb{R}^{n})$. Note
that Moyal's identity is equivalent to the equality:
\begin{equation}
(\psi_{1}|\psi_{2})\overline{(\phi_{1}|\phi_{2})}=(\iint\nolimits_{\mathbb{R}%
^{2n}}\langle f,T(z)\psi_{1}\rangle T(z)\psi_{2}dz|\psi_{2}).
\end{equation}
Setting $\psi_{1}=\psi$ and assuming that $(\phi_{1}|\phi_{2})\neq0$ this
equality becomes a resolution of the identity:
\begin{equation}
\psi=(\phi_{1}|\phi_{2})^{-1}\iint_{\mathbb{R}^{2n}}(\psi|T(z)\phi
_{1})T(z)\phi_{2}dz.
\end{equation}
In the language of frames in Hilbert spaces, this resolution of the identity
amounts to the statement that the set $\{T(z)\psi:z\in\mathbb{R}^{2n}\}$ is a
{tight frame} for $L^{2}(\mathbb{R}^{n})$ (see \cite{gr00} for a thorough
discussion of frames in time-frequency analysis). In the present setting we
can always find $\phi_{1}$ in $M_{v_{s}}^{1}(\mathbb{R}^{n})$ or in
$\mathcal{S}(\mathbb{R}^{n})$, i.e. there exist tight frames $\{T(z)\psi
_{1}:z\in\mathbb{R}^{2n}\}$ for $L^{2}(\mathbb{R}^{n})$ with good phase space
localization. The main purpose of this section is to discuss the consequences
of the square-integrability of $\tilde{T}(z)$ on $L^{2}(\mathbb{R}^{2n})$.

Unless otherwise specified $\phi$ will denote a function in $\mathcal{S}%
(\mathbb{R}^{n})$ such that $||\phi||=1$; we will call $\phi$ a
\textquotedblleft window\textquotedblright.

\subsection{Definition and functional properties}

By definition the wavepacket transform $\mathcal{U}_{\phi}$ on $L^{2}%
(\mathbb{R}^{n})$ with window $\phi$ is defined by%
\begin{equation}
\mathcal{U}_{\phi}\psi(z)=\left(  \tfrac{\pi\hbar}{2}\right)  ^{n/2}%
W(\psi,\phi)(\tfrac{1}{2}z)\text{ , }\psi\in\mathcal{S}^{\prime}%
(\mathbb{R}^{n}); \label{wpt}%
\end{equation}
here $W(\psi,\phi)$ is the cross-Wigner distribution, defined for\textit{
}$\psi,\phi\in L^{2}(\mathbb{R}^{n})$ by\textit{ }
\begin{equation}
W(\psi,\phi)(z)=\left(  \tfrac{1}{2\pi\hbar}\right)  ^{n}\int_{\mathbb{R}^{n}%
}e^{-\tfrac{i}{\hbar}y\cdot\eta}\psi(x+\tfrac{1}{2}\eta)\overline
{\phi(x-\tfrac{1}{2}\eta)}d\eta. \label{wigner}%
\end{equation}
We observe for further use that $\mathcal{U}_{\phi}\psi$ can be written%
\begin{equation}
\mathcal{U}_{\phi}\psi(z)=\left(  \tfrac{1}{2\pi\hbar}\right)  ^{n/2}%
(\widehat{\Pi}(\tfrac{1}{2}z)\psi|\phi) \label{ufgr}%
\end{equation}
where $\widehat{\Pi}(z_{0})$ is the Grossmann--Royer operator
\cite{Grossmann,Royer} defined by
\begin{equation}
\widehat{\Pi}(z_{0})\psi(x)=e^{\frac{2i}{\hbar}y_{0}(x-x_{0})}\psi(2x_{0}-x).
\label{GR}%
\end{equation}

It is useful to have a result showing how the windowed wavepacket transform
behaves under the action of symplectic linear automorphisms.

\begin{proposition}
\label{symwpt}Let $s\in\operatorname*{Sp}(2n,\mathbb{R})$ and $\psi\in
L^{2}(\mathbb{R}^{n})$. We have
\begin{equation}
\mathcal{U}_{\phi}\psi(s^{-1}z)=\mathcal{U}_{S\phi}(S\psi)(z) \label{symco}%
\end{equation}
where $S$ is any of the two operators in the metaplectic group
$\operatorname*{Mp}(2n,\mathbb{R})$ covering $s$.
\end{proposition}

\begin{proof}
It immediately follows from the well-known covariance formula
\[
W(\psi,\phi)\circ s^{-1}=W(S\psi,S\phi)
\]
satisfied by the cross-Wigner distribution (see for instance de Gosson
\cite{Birk}, Proposition 7.14, p.207).
\end{proof}

The following theorem, part of which was proven in \cite{Birk,RMP}, summarizes
the main functional analytical properties of the wavepacket transform.

\begin{theorem}
\label{th1}The wavepacket transform $\mathcal{U}_{\phi}$ is a partial isometry
from $L^{2}(\mathbb{R}^{n})$ into $L^{2}(\mathbb{R}^{2n})$. More explicitly,
the wavepacket transform has the following properties: (i) $\mathcal{U}_{\phi
}$ is a linear isometry of $L^{2}(\mathbb{R}^{n})$ onto a closed subspace
$\mathcal{H}_{\phi}$ of $L^{2}(\mathbb{R}^{2n})$; (ii) Let $\mathcal{U}_{\phi
}^{\ast}:L^{2}(\mathbb{R}^{2n})\longrightarrow L^{2}(\mathbb{R}^{n})$ be the
adjoint of $\mathcal{U}_{\phi}$. We have $\mathcal{U}_{\phi}^{\ast}%
\mathcal{U}_{\phi}=I$ on $L^{2}(\mathbb{R}^{n})$ and the operator $P_{\phi
}=\mathcal{U}_{\phi}\mathcal{U}_{\phi}^{\ast}$ is the orthogonal projection in
$L^{2}(\mathbb{R}^{2n})$ onto the space $\mathcal{H}_{\phi}$; (iii) The
inverse $\mathcal{U}_{\phi}^{-1}:\mathcal{H}_{\phi}\longrightarrow
L^{2}(\mathbb{R}^{n})$ is given by the formula%
\begin{equation}
\psi(x)=\frac{(2\pi\hbar)^{n/2}}{(\gamma|\phi)}\int_{\mathbb{R}^{n}%
}\mathcal{U}_{\phi}\psi(z_{0})\widehat{\Pi}(\tfrac{1}{2}z_{0})\gamma(x)dz_{0}
\label{cg}%
\end{equation}
where $\gamma\in L^{2}(\mathbb{R}^{n})$ is such that $(\gamma|\phi)\neq0$;
(iv) The adjoint $\mathcal{U}_{\phi}^{\ast}$ of $\mathcal{U}_{\phi}$ is given
by%
\begin{equation}
\mathcal{U}_{\phi}^{\ast}\Psi(z)=\left(  \tfrac{2}{\pi\hbar}\right)
^{n/2}\int_{\mathbb{R}^{n}\times\mathbb{R}^{n}}e^{\frac{2i}{\hbar}p\cdot
(x-y)}\phi(2y-x)\Psi(y,p)dpdy. \label{uadj}%
\end{equation}

\end{theorem}

\begin{proof}
Properties (i) and (ii) were proven in \cite{RMP} and \cite{Birk}, Chapter 10,
\S 2; note that the fact that $\mathcal{U}_{\phi}$ is an isometry immediately
follows from Moyal's identity \eqref{Moyal}:%
\begin{equation}
((W(\psi,\phi)|W(\psi^{\prime},\phi^{\prime})))=\left(  \tfrac{1}{2\pi\hbar
}\right)  ^{n}(\psi|\psi^{\prime})\overline{(\phi|\phi^{\prime})}.
\label{moyal}%
\end{equation}
Let us prove the inversion formula (\ref{cg}). Set%
\[
\psi^{\prime}(x)=C_{\gamma}\int_{\mathbb{R}^{n}}\Psi(z_{0})\widehat{\Pi
}(\tfrac{1}{2}z_{0})\gamma(x)dz_{0}%
\]
where $C_{\gamma}$ is a constant. For every $\theta\in L^{2}(\mathbb{R}^{n})$
we have, using successively (\ref{ufgr}) and the Moyal identity (\ref{moyal}%
),
\begin{align*}
(\psi^{\prime}|\theta)  &  =C_{\gamma}\int_{\mathbb{R}^{n}}\mathcal{U}_{\phi
}\psi(z_{0})(\widehat{\Pi}(\tfrac{1}{2}z_{0})\gamma|\theta)dz_{0}\\
&  =C_{\gamma}2^{-n/2}\left(  \pi\hbar\right)  ^{3n/2}\int_{\mathbb{R}^{2n}%
}W(\psi,\phi)(\tfrac{1}{2}z_{0})W(\gamma,\theta)(\tfrac{1}{2}z_{0})dz_{0}\\
&  =C_{\gamma}\left(  2\pi\hbar\right)  ^{3n/2}\int_{\mathbb{R}^{2n}}%
W(\psi,\phi)(z_{0})W(\gamma,\theta)(z_{0})dz_{0}\\
&  =C_{\gamma}\left(  2\pi\hbar\right)  ^{n/2}(\psi|\theta)\overline
{(\phi|\gamma)}.
\end{align*}
It follows that $\psi^{\prime}=\psi$ if we choose the constant $C_{\gamma}$ so
that $C_{\gamma}\left(  2\pi\hbar\right)  ^{n/2}\overline{(\phi|\gamma)}=1$,
which proves (\ref{cg}). Formula (\ref{uadj}) for $\mathcal{U}_{\phi}^{\ast}$
is obtained by a straightforward calculation using the identity $(\mathcal{U}%
_{\phi}\psi|\Psi)_{L^{2}(\mathbb{R}^{2n})}=(\psi|\mathcal{U}_{\phi}^{\ast}%
\Psi)_{L^{2}(\mathbb{R}^{n})}$ and the definition of $\mathcal{U}_{\phi}$ in
terms of the cross-Wigner distribution.
\end{proof}

\subsection{The intertwining property}

Here is the key result which shows how the operators $A$ and $\widetilde{A}$
are linked by the wavepacket transforms:

\begin{proposition}
\label{inter}Let $\mathcal{U}_{\phi}$ be an arbitrary wavepacket transform.
The following intertwining formula hold:
\begin{align}
\widetilde{T}(z)\mathcal{U}_{\phi}  &  =\mathcal{U}_{\phi}T(z)\text{ \ and
\ }\widetilde{A}\mathcal{U}_{\phi}=\mathcal{U}_{\phi}A\text{\ }
\label{intertw}\\
\mathcal{U}_{\phi}^{\ast}\widetilde{T}(z)\mathcal{U}_{\phi}  &  =T(z)\text{
\ and \ }\mathcal{U}_{\phi}^{\ast}\widetilde{A}\mathcal{U}_{\phi}=A.
\label{intertx}%
\end{align}

\end{proposition}

\begin{proof}
The proof of formulae (\ref{intertw}) is purely computational (see
\cite{Birk}, Theorem 10.10, p.317, where $\widetilde{T}$ and $A$ are denoted
there by $T_{\text{ph}}$ and $A_{\text{ph}}$, respectively). Formulae
(\ref{intertx}) immediately follow since $\mathcal{U}_{\phi}^{\ast}%
\mathcal{U}_{\phi}=I$ on $L^{2}(\mathbb{R}^{n})$.
\end{proof}

For instance, if
\[
H_{\text{har}}=-\frac{\hbar^{2}}{2m}\frac{\partial^{2}}{\partial x^{2}}%
+\frac{m\omega^{2}}{2}x^{2}%
\]
is the harmonic oscillator operator and
\[
H_{\text{sym}}=-\frac{\hbar^{2}}{2m}\Delta_{x,y}-i\hbar\omega_{L}\left(
y\frac{\partial}{\partial x}-x\frac{\partial}{\partial y}\right)
+\frac{m\omega_{L}^{2}}{2}(x^{2}+y^{2})
\]
is the magnetic operator considered in the introduction we have
\begin{equation}
H_{\text{sym}}\mathcal{U}_{\phi}=\mathcal{U}_{\phi}H_{\text{har}}\text{\ }
\label{huh}%
\end{equation}
as an immediate consequence of the second formula (\ref{intertw}). We will use
this intertwining relation in Section \ref{sec3} to recover the Landau levels
and the corresponding Landau eigenfunctions.

\subsection{WPT and modulation spaces}

The \textit{cross-Wigner transform} (\ref{wigner}) is related to the STFT by
the formula
\begin{equation}
W(\psi,\phi)(z)=\left(  \tfrac{2}{\pi\hbar}\right)  ^{n/2}e^{\frac{2i}{\hbar
}y\cdot x}V_{\phi_{\sqrt{2\pi\hbar}}^{\vee}}\psi_{\sqrt{2\pi\hbar}}\left(
\tfrac{1}{\sqrt{2\pi\hbar}}z\right)  \label{stuff}%
\end{equation}
where we set as usual $\psi_{\lambda}(x)=\psi(\lambda x)$ for a non-zero
$\lambda\in\mathbb{R}$, $\phi^{\vee}(x)=\phi(-x)$. It follows that there is a
simple relationship between the windowed wavepacket transform $\mathcal{U}%
_{\phi}$ and the short-time Fourier transform $V_{\phi}$. In fact, using
formula (\ref{stuff}) relating $V_{\phi}$ to the cross-Wigner transform
together with definition (\ref{wpt}) of $\mathcal{U}_{\phi}$ we have:%
\begin{equation}
\mathcal{U}_{\phi}\psi(z)=e^{\frac{i}{2\hbar}y\cdot x}V_{\phi_{\sqrt{2\pi
\hbar}}^{\vee}}\psi_{\sqrt{2\pi\hbar}}\left(  z/2\sqrt{2\pi\hbar}\right)
\label{uv}%
\end{equation}
and hence also%
\begin{equation}
V_{\phi}\psi(z)=e^{-4\pi iy\cdot x}\mathcal{U}_{\phi_{1/\sqrt{2\pi\hbar}%
}^{\vee}}\psi_{1/\sqrt{2\pi\hbar}}\left(  2\sqrt{2\pi\hbar}z\right)  .
\label{v u}%
\end{equation}

\begin{proposition}
\label{gros}(i) We have $\psi\in M_{v_{s}}^{1}(\mathbb{R}^{n})$ if and only if
$\mathcal{U}_{\phi}\psi\in L_{v_{s}}^{1}(\mathbb{R}^{2n})$ for one (and hence
for all) window(s) $\phi\in\mathcal{S}(\mathbb{R}^{n})$. (ii) For $\phi
\in\mathcal{S}(\mathbb{R}^{n})$, $\phi\neq0$, the formula%
\begin{equation}
||\psi||_{M_{v_{s}}^{1},\phi}=|||\mathcal{U}_{\phi}\psi|||_{L_{v_{s}}^{1}}
\label{ms}%
\end{equation}
defines a family of norms on $M_{v_{s}}^{1}(\mathbb{R}^{n})$ which are
equivalent to the norms $||\psi||_{M_{v_{s}}^{1}}^{\phi}$ defined by
(\ref{norm1s}). (iii) The operator $\mathcal{U}_{\phi}^{\ast}$ maps $L_{v_{s}%
}^{1}(\mathbb{R}^{2n})$ into $M_{v_{s}}^{1}(\mathbb{R}^{n})$ and the inversion
formula (\ref{cg}) in Theorem \ref{th1} holds in $M_{v_{s}}^{1}(\mathbb{R}%
^{n})$. (iv) $M_{v_{s}}^{1}(\mathbb{R}^{n})$ is invariant under the action of
the metaplectic group $\operatorname*{Mp}(2n,\mathbb{R})$.
\end{proposition}

\begin{proof}
(i) Immediately follows from formula (\ref{uv}) using Lemma \ref{lemsca}. The
statement (ii) follows from Proposition 11.3.2 in \cite{gr00}; (iii) follows
from Corollary 11.3.4 in \cite{gr00}. Properties (iv) and (v) have been
established in Proposition 11.3.2 of \cite{gr00}.
\end{proof}

\section{Spectral Properties\label{sec3}}

We are going to use the results above to compare the spectral properties of
$A$ and $\widetilde{A}$. We assume throughout that the operators $A$ and
$\widetilde{A}$ are defined on some dense subspace of $L^{2}(\mathbb{R}^{n})$
and $L^{2}(\mathbb{R}^{2n})$, respectively.

\subsection{General results}

The following result is very useful for the study of the eigenvectors of the
LW operators:

\begin{lemma}
\label{Wong}Let $(\phi_{j})_{j\in F}$ be an arbitrary orthonormal basis of
$L^{2}(\mathbb{R}^{n})$; setting $\Phi_{j,k}=\mathcal{U}_{\phi_{j}}\phi_{k}$
the family $\{\Phi_{j,k}:(j,k)\in F\times F\}$ forms an orthonormal basis of
$L^{2}(\mathbb{R}^{2n})$, i.e. $L^{2}(\mathbb{R}^{2n})=\bigoplus
\nolimits_{j}\mathcal{H}_{\phi_{j}}$ (Hilbert sum).
\end{lemma}

\begin{proof}
It is sufficient to prove the result for $\mathcal{U}_{\phi_{j}}$. Since the
$\mathcal{U}_{\phi_{j}}$ are isometries the vectors $\Phi_{j,k}$ form an
orthonormal system. Let us show that if $\Psi\in L^{2}(\mathbb{R}^{2n})$ is
orthogonal to the family $(\Phi_{j,k})_{j,k}$ (and hence to all the spaces
$\mathcal{H}_{\phi_{j}}$) then it is the zero vector; it will follow that
$(\Phi_{j,k})_{j,k}$ is a basis. Assume that $((\Psi|\Phi_{jk}))=0$ for all
$j,k.$ Since we have
\[
((\Psi|\Phi_{jk}))=((\Psi|\mathcal{U}_{\phi_{j}}\phi_{k}))=(\mathcal{U}%
_{\phi_{j}}^{\ast}\Psi|\phi_{k})
\]
this means that $\mathcal{U}_{\phi_{j}}^{\ast}\Psi=0$ for all $j$ since
$(\phi_{j})_{j}$ is a basis. In view of Theorem \ref{th1}(ii) we thus have
$P_{\phi_{j}}\Psi=0$ for all $j$ so that $\Psi$ is orthogonal to all
$\mathcal{H}_{\phi_{j}}.$
\end{proof}

\begin{theorem}
\label{eigen1}(i) The eigenvalues of the operators $A$ and $\widetilde{A}$ are
the same; (ii) Let $\psi$ be an eigenvector of $A$: $A\psi=\lambda\psi$. Then
$\Psi=\mathcal{U}_{\phi}\psi$ is an eigenvector of $A$ corresponding to the
same eigenvalue: $\widetilde{A}\Psi=\lambda\Psi$. (ii) Conversely, if $\Psi$
is an eigenvector of $A$ then $\psi=\mathcal{U}_{\phi}^{\ast}\Psi$ is an
eigenvector of $A$ corresponding to the same eigenvalue.
\end{theorem}

\begin{proof}
(i) That every eigenvalue of $A$ also is an eigenvalue of $\widetilde{A}$ is
clear: if $A\psi=\lambda\psi$ for some $\psi\neq0$ then
\[
\widetilde{A}(\mathcal{U}_{\phi}\psi)=\mathcal{U}_{\phi}A\psi=\lambda
\mathcal{U}_{\phi}\psi
\]
and $\Psi=\mathcal{U}_{\phi}\psi\neq0$ ; this proves at the same time that
$\mathcal{U}_{\phi}\psi$ is an eigenvector of $A$ because $\mathcal{U}_{\phi}$
has kernel $\{0\}$. (ii) Assume conversely that $\widetilde{A}\Psi=\lambda
\Psi$ for $\Psi\in L^{2}(\mathbb{R}^{2n})$, $\Psi\neq0$, and $\lambda
\in\mathbb{R}$. For every $\phi$ we have
\[
A\mathcal{U}_{\phi}^{\ast}\Psi=\mathcal{U}_{\phi}^{\ast}\widetilde{A}%
\Psi=\lambda\mathcal{U}_{\phi}^{\ast}\Psi
\]
hence $\lambda$ is an eigenvalue of $A$ and $\psi$ an eigenvector if
$\psi=\mathcal{U}_{\phi}^{\ast}\Psi\neq0$. We have $\mathcal{U}_{\phi}%
\psi=\mathcal{U}_{\phi}\mathcal{U}_{\phi}^{\ast}\Psi=P_{\phi}\Psi$ where
$P_{\phi}$ is the orthogonal projection on the range $\mathcal{H}_{\phi}$ of
$\mathcal{U}_{\phi}$. Assume that $\psi=0$; then $P_{\phi}\Psi=0$ for every
$\phi\in\mathcal{S}(\mathbb{R}^{n}),$ and hence $\Psi=0$ in view of Lemma
\ref{Wong}.
\end{proof}

The reader is urged to remark that the result above is quite general: it doses
not make any particular assumptions on the operator $A$ (in particular it is
not assumed that $A$ is self-adjoint), and the multiplicity of the eigenvalues
can be arbitrary.

Let us specialize the results above to the case where $\widetilde{A}$ is
(essentially) self-adjoint:

\begin{corollary}
\label{kernel}Suppose that $A$ is a self-adjoint operator on $L^{2}%
(\mathbb{R}^{n})$ and that each of the eigenvalues $\lambda_{0},\lambda
_{1},...,\lambda_{j},...$ has multiplicity one. Let $\psi_{0},\psi
_{1},...,\psi_{j},...$ be a corresponding sequence of orthonormal
eigenvectors. Let $\Psi_{j}$ be an eigenvector of $\widetilde{H}$
corresponding to the eigenvalue $\lambda_{j}$. There exists a sequence
$(\alpha_{j,k})_{k}$ of complex numbers such that
\begin{equation}
\Psi_{j}=\sum_{\ell}\alpha_{j,\ell}\Psi_{j,\ell}\text{ \ with \ }\Psi_{j,\ell
}=\mathcal{U}_{\psi_{\ell}}\psi_{j}\in\mathcal{H}_{j}\cap\mathcal{H}_{\ell
}\text{.} \label{fifi}%
\end{equation}

\end{corollary}

\begin{proof}
We know from Theorem \ref{eigen1} above that $A$ and $\widetilde{A}$ have same
eigenvalues and that $\Psi_{j,k}=W_{\psi_{k}}\psi_{j}$ satisfies
$\widetilde{H}\Psi_{j,k}=\lambda_{j}\Psi_{j,k}$. Since $A$ is self-adjoint its
eigenvectors $\psi_{j}$ form an orthonormal basis of $L^{2}(\mathbb{R}^{n})$;
it follows from Lemma \ref{Wong} that the $\Psi_{j,k}$ form an orthonormal
basis of $L^{2}(\mathbb{R}^{2n})$, hence there exist non-zero scalars
$\alpha_{j,k,\ell}$ such that $\Psi_{j}=\sum_{k,\ell}\alpha_{j,k,\ell}%
\Psi_{k,\ell}$. We have, by linearity and using the fact that $\widetilde
{A}\Psi_{k,\ell}=\lambda_{k}\Psi_{k,\ell}$,
\[
\widetilde{A}\Psi_{j}=\sum_{k,\ell}\alpha_{j,k,\ell}\widetilde{A}\Psi_{k,\ell
}=\sum_{k,\ell}\alpha_{j,k,\ell}\lambda_{k}\Psi_{k,\ell}.
\]
On the other hand we also have $\widetilde{A}\Psi_{j}=\lambda_{j}\Psi_{j}$,
\[
\widetilde{H}\Psi_{j}=\lambda_{j}\Psi_{j}=\sum_{j,k}\alpha_{j,k,\ell}%
\lambda_{j}\Psi_{k,\ell}%
\]
and this is only possible if $\alpha_{j,k,\ell}=0$ for $k\neq j$; setting
$\alpha_{j,\ell}=\alpha_{j,j,\ell}$ formula (\ref{fifi}) follows. (That
$\Psi_{j,\ell}\in\mathcal{H}_{j}\cap\mathcal{H}_{\ell}$ is clear using the
definition of $\mathcal{H}_{\ell}$ and the sesquilinearity of the cross-Wigner transform.)
\end{proof}

\subsection{Shubin classes}

Shubin has introduced in \cite{sh87} very convenient symbol classes for
studying global hypoellipticity. These \textquotedblleft Shubin
classes\textquotedblright\ are defined as follows: let $H\Gamma_{\rho}%
^{m_{1},m_{0}}(\mathbb{R}^{2n})$ ($m_{0},m_{1}\in\mathbb{R}$ and $0<\rho\leq
1$) be the complex vector space of all functions $a\in C^{\infty}%
(\mathbb{R}^{2n})$ for which there exists a number $R\geq0$ such that for
$|z|\geq R$ we have
\begin{equation}
C_{0}|z|^{m_{0}}\leq|a(z)|\leq C_{1}|z|^{m_{1}}\text{ \ , \ }|\partial
_{z}^{\alpha}a(z)|\leq C_{\alpha}|a(z)||z|^{-\rho|\alpha|} \label{shu}%
\end{equation}
for some constants $C_{0},C_{1},C_{\alpha}\geq0$; we are using here
multi-index notation $\alpha=(\alpha_{1},...,\alpha_{2n})\in\mathbb{N}^{n}$,
$|\alpha|=\alpha_{1}+\cdot\cdot\cdot+\alpha_{2n}$, $\ $and $\partial
_{z}^{\alpha}=\partial_{x_{1}}^{\alpha_{1}}\cdot\cdot\cdot\partial_{x_{n}%
}^{\alpha_{n}}\partial_{y_{1}}^{\alpha_{n+1}}\cdot\cdot\cdot\partial_{y_{n}%
}^{\alpha_{2n}}.$ We notice that the Shubin classes are invariant under linear
changes of variables: if $f\in GL(2n,\mathbb{R})$ and $a\in H\Gamma_{\rho
}^{m_{1},m_{0}}(\mathbb{R}^{2n})$ then $a\circ f\in H\Gamma_{\rho}%
^{m_{1},m_{0}}(\mathbb{R}^{2n})$. In particular they are invariant under
linear symplectic transformations.

We denote by $HG_{\rho}^{m_{1},m_{0}}(\mathbb{R}^{n})$ the class of operators
$A$ with $\tau$-symbols $a_{\tau}$ belonging to $H\Gamma_{\rho}^{m_{1},m_{0}%
}(\mathbb{R}^{2n})$; this means that for every $\tau\in\mathbb{R}$ there
exists $a_{\tau}\in H\Gamma_{\rho}^{m_{1},m_{0}}(\mathbb{R}^{2n})$ such that
\[
Au(x)=\left(  \tfrac{1}{2\pi}\right)  ^{n}\iint\nolimits_{\mathbb{R}^{2n}%
}e^{i(x-y)\cdot\xi}a_{\tau}((1-\tau)x+\tau y,\xi)u(y)dyd\xi;
\]
choosing $\tau=\frac{1}{2}$ this means, in particular, that every operator
with Weyl symbol $a\in H\Gamma_{\rho}^{m_{1},m_{0}}(\mathbb{R}^{2n})$ is in
$HG_{\rho}^{m_{1},m_{0}}(\mathbb{R}^{2n})$. Conversely, the condition $a\in
H\Gamma_{\rho}^{m_{1},m_{0}}(\mathbb{R}^{2n})$ is also sufficient, because if
$a_{\tau}\in H\Gamma_{\rho}^{m_{1},m_{0}}(\mathbb{R}^{2n})$ is true for some
$\tau$ then it is true for all $\tau$.

Shubin \cite{sh87} (Chapter 4) proves the following result:

\begin{proposition}
[Shubin]Let $A\in HG_{\rho}^{m_{1},m_{0}}(\mathbb{R}^{2n})$ with $m_{0}>0$. If
$A$ is formally self-adjoint, that is if $(A\psi|\phi)=(\psi|A\phi)$ for all
$\psi,\phi\in C_{0}^{\infty}(\mathbb{R}^{n})$, then: (i) $A$ is essentially
self-adjoint and has discrete spectrum in $L^{2}(\mathbb{R}^{n})$; (ii) There
exists an orthonormal basis of eigenfunctions $\phi_{j}\in\mathcal{S}%
(\mathbb{R}^{n})$ ($j=1,2,...$) with eigenvalues $\lambda_{j}\in\mathbb{R}$
such that $\lim_{j\rightarrow\infty}|\lambda_{j}|=\infty$.
\end{proposition}

This result has the following consequence for LW operators:

\begin{corollary}
Let $A\in HG_{\rho}^{m_{1},m_{0}}(\mathbb{R}^{2n})$ be formally self-adjoint.
Then the LW operator $\widetilde{A}$ has discrete spectrum $(\lambda
_{j})_{j\in\mathbb{N}}$ and $\lim_{j\rightarrow\infty}|\lambda_{j}|=\infty$
and the eigenfunctions of $\widetilde{A}$ are in this case given by $\Phi
_{jk}=\mathcal{U}_{\phi_{j}}\phi_{k}$ where the $\phi_{j}$ are the
eigenfunctions of $A$; (iv) We have $\Phi_{jk}\in\mathcal{S}(\mathbb{R}^{2n})$
and the $\Phi_{jk}$ form an orthonormal basis of $\phi_{j}\in\mathcal{S}%
(\mathbb{R}^{n})$.
\end{corollary}

\begin{proof}
It is an immediate consequence of Theorem \ref{eigen1} using the proposition above.
\end{proof}

\subsection{Gelfand triples}

Dirac already emphasized in his fundamental work \cite{Dirac} the relevance of
\textit{rigged Hilbert spaces} for quantum mechanics. Later Schwartz provided
an instance of rigged Hilbert spaces based on his class of test functions and
on tempered distributions. Later Gelfand and Shilov formalized the
construction of Schwartz and Dirac and introduced what is nowadays known as
\textit{Gelfand triples} The prototypical example of a Gelfand triple is
$(\mathcal{S}(\mathbb{R}^{n},L^{2}(\mathbb{R}^{n},\mathcal{S}^{\prime
}(\mathbb{R}^{n}))$. In the last decade Feichtinger and some of his
collaborators (see \cite{feko98,dofegr06,cofelu07}) emphasized the relevance
of the Gelfand triple $(M_{v_{s}}^{1}(\mathbb{R}^{n}),L^{2}(\mathbb{R}%
^{n}),M_{1/v_{s}}^{\infty}(\mathbb{R}^{n}))$ in time-frequency analysis. An
important feature of Gelfand triples is the existence of a kernel theorem, as
we explained in Subsection \ref{subseckern}. In the present investigation
these classes of Gelfand triples will allow us to treat the case of the
continuous spectrum of selfadjoint operators.

The main idea underlying the notion of Gelfand triple is the observation, that
a triple of spaces -- consisting of the Hilbert space itself, a small
(topological vector) space contained in the Hilbert space, and its dual --
allows a much better description of the spectrum. The main appeal of the
notion of Banach triple is, in our context, the fact that we can even take a
Banach space, namely the modulation space $M_{v_{s}}^{1}(\mathbb{R}^{n})$.

\begin{definition}
A (Banach) \textit{Gelfand triple} $(\mathcal{B},\mathcal{H},\mathcal{B}%
^{\prime})$ consists of a Banach space $\mathcal{B}$ which is continuously and
densely embedded into a Hilbert space ${\mathcal{H}}$, which in turn is
$w^{\ast}$-continuously and densely embedded into the dual Banach space
$\mathcal{B}^{\prime}$.
\end{definition}

In this setting the inner product on $\mathcal{H}$ extends in a natural way to
a pairing between $\mathcal{B}$ and $\mathcal{B}^{\prime}$ producing an
anti-linear functional $F$ of the same norm. The framework of the Gelfand
triple $(\mathcal{S}(\mathbb{R}^{n},L^{2}(\mathbb{R}^{n}),\mathcal{S}^{\prime
}(\mathbb{R}^{n}))$ or more generally of $(M_{v_{s}}^{1}(\mathbb{R}^{n}%
),L^{2}(\mathbb{R}^{n}),M_{1/v_{s}}^{\infty}(\mathbb{R}^{n}))$ allows one to
formulate a spectral theorem for selfadjoint operators on $\mathcal{S}%
(\mathbb{R}^{n})$ or $M_{v_{s}}^{1}(\mathbb{R}^{n})$. If $F(A\psi)=\lambda
F(\psi)$ holds for all $\psi\in M_{v_{s}}^{1}(\mathbb{R}^{n})$ or in
$\mathcal{S}(\mathbb{R}^{n})$ in the distributional sense, then $\lambda$ is
called a \textit{generalized eigenvalue} to the \textit{generalized
eigenvector} of the selfadjoint operator $A$. For a given generalized
eigenvalue $\lambda\in\mathbb{C}$ we denote by $E_{\lambda}$ be the set of all
generalized eigenvectors $F$ in $M_{1/v_{s}}^{\infty}(\mathbb{R}^{n})$ or
$\mathcal{S}(\mathbb{R}^{n})$, respectively. The set of all generalized
eigenvalues $\cup_{\lambda}E_{\lambda}$ is called \textit{complete}, if for
any $\psi,\phi$ in $M_{1/v_{s}}^{\infty}(\mathbb{R}^{n})$ or $\mathcal{S}%
(\mathbb{R}^{n})$ such that $F(\psi)=F(\phi)$ for all $F\in\cup_{\lambda
}E_{\lambda}$, then $\psi=\phi$.

\begin{theorem}
Let $T$ be a selfadjoint operator on $M_{v_{s}}^{1}(\mathbb{R}^{n})$ or
$\mathcal{S}(\mathbb{R}^{n})$. Then all generalized eigenvalues $\lambda$ are
real numbers and $L^{2}(\mathbb{R}^{n})$ can be written as a direct sum of
Hilbert spaces $\mathcal{H}(\lambda)$ such that $E_{\lambda}\subset
\mathcal{H}_{\lambda}$, and such that the $\lambda$-component of $Tf$ is given
by $(Af)_{\lambda}=\lambda f$ for all $M_{v_{s}}^{1}(\mathbb{R}^{n})$ or
$\mathcal{S}(\mathbb{R}^{n})$. Moreover, the set of generalized eigenvectors
$\cup_{\lambda}E_{\lambda}$ is complete.
\end{theorem}

As an illustration we treat generalized eigenvectors of the translation
operator $T_{x}f(y)=f(y-x)$. We interpret the characters $\chi_{\omega
}(x)=e^{-2\pi i\omega\cdot x}$ as generalized eigenvectors for the translation
operator $T_{x}$ on $M_{v_{s}}^{1}(\mathbb{R}^{n})$. Furthermore the set of
generalized eigenvectors $\{\chi_{\omega}:\omega\in\mathbb{R}^{n}\}$ is
complete by Plancherel's theorem, i.e., if the Fourier transform $\hat
{f}(\omega)=\langle\chi_{\omega},f\rangle$ vanishes for all $\omega
\in\mathbb{R}^{n}$ implies $f\equiv0$. This suggests to think of the Fourier
transform of $f$ at frequency $\omega$ as the evaluation of the linear
functional $\langle\chi_{\omega},f\rangle$.

Therefore the preceding theorem allows us to deal with the continuous spectrum
as treated in Theorem \ref{eigen1} for the discrete spectrum.

\subsection{Application to the Landau levels}

As an illustration consider the harmonic oscillator Hamiltonian (\ref{har}) of
the Introduction. To simplify notation we take $m=\omega=1$ (this corresponds
to the choice $\gamma=\mu=1$ for $\widetilde{A}^{\gamma,\mu}$); in addition we
choose units in which $\hbar=1$ In view of the results above the spectra of
the harmonic oscillator Hamiltonian (\ref{har}) and of the magnetic operator
(\ref{magn1}) are identical. The eigenvalues of the first are the numbers
$\lambda_{k}=k+\frac{1}{2}$ ($k$ an integer). These are the well-known Landau
energy levels \cite{lala97}. The harmonic oscillator operator (\ref{har})
satisfies the assumptions of Corollary \ref{kernel}. The normalized
eigenvectors are the rescaled Hermite functions%
\[
\phi_{k}(x)=(2^{k}k!\sqrt{\pi})^{-\frac{1}{2}}e^{-\frac{1}{2}x^{2}}%
\mathcal{H}_{k}(x).
\]
where
\[
\mathcal{H}_{k}(x)=(-1)^{km}e^{x^{2}}\left(  \tfrac{d}{dx}\right)
^{k}e^{-x^{2}}%
\]
is the $k$-th Hermite polynomial. Using definition (\ref{wpt}) of the
wavepacket transform together with known formulae for the cross-Wigner
transform of Hermite functions (Thangavelu \cite{th93}, Chapter 1, Wong
\cite{Wong}, Chapter 24, Theorem 24.1) one finds that the eigenvectors of the
magnetic operator are linear superpositions of the functions%
\[
\Phi_{j+k,k}(z)=(-1)^{j}\tfrac{1}{\sqrt{2\pi}}\left(  \tfrac{j!}%
{(j+k)!}\right)  ^{\frac{1}{2}}2^{-\frac{k}{2}}z^{k}\mathcal{L}_{j}^{k}%
(\tfrac{1}{2}|z|^{2})e^{-\frac{|z|^{2}}{4}}%
\]
and $\Phi_{j,j+k}=\overline{\Phi_{j+k,k}}$ for $k=0,1,2,...$; in the
right-hand side $z$ is interpreted as $x+iy$ and%
\[
\mathcal{L}_{j}^{k}(x)=\tfrac{1}{j!}x^{-k}e^{x}\left(  \tfrac{d}{dx}\right)
^{j}(e^{-x}x^{j+k})\text{ , }x>0
\]
is the Laguerre polynomial of degree $j\ $and order $k$. In particular we
recover the textbook result that the eigenspace of the ground state is spanned
by the functions%
\[
\Phi_{0,k}(x,y)=(k!2^{k+1}\pi)^{-1/2}(x-iy)^{k}e^{-\frac{1}{4}(x^{2}+y^{2})};
\]
notice that this eigenspace is just $\mathcal{H}_{\phi_{0}}.$ Finally we want
to mention that the intertwining between the Weyl calculus and the
Landau--Weyl calculus allows one to define annihilation and creation operators
as in the case of the harmonic oscillator. Therefore our calculus provides us
with natural operators that allow us to \textquotedblleft
move\textquotedblright\ between the eigenvectors of the Landau levels. We will
come back to this issue in a forthcoming work.

\section{Regularity and Hypoellipticity Results\label{sec4}}

We begin by stating a few boundedness results for Weyl and Landau--Weyl
operators in modulation spaces. The main result of this section is Theorem
\ref{aglo} where we prove a global hypoellipticity result for Landau--Weyl
operators whose symbol belong to the Shubin class $H\Gamma_{\rho}^{m_{1}%
,m_{0}}(\mathbb{R}^{2n})$.

\subsection{Global hypoellipticity}

In \cite{sh87} (Corollary 25.1, p. 186) Shubin has introduced the notion of
global hypoellipticity (also see Boggiatto et al. \cite{boburo96}, p. 70).
This notion is more useful in quantum mechanics than the usual hypoellipticity
because it incorporates the decay at infinity of the involved distributions.

\begin{definition}
\label{defglob}We will say that a linear operator $A:\mathcal{S}^{\prime
}(\mathbb{R}^{n})\longrightarrow\mathcal{S}^{\prime}(\mathbb{R}^{n})$ is
\textquotedblleft globally hypoelliptic\textquotedblright\ if we have%
\begin{equation}
\psi\in\mathcal{S}^{\prime}(\mathbb{R}^{n})\text{ and }A\psi\in\mathcal{S}%
(\mathbb{R}^{n})\Longrightarrow\psi\in\mathcal{S}(\mathbb{R}^{n}).
\label{glob1}%
\end{equation}

\end{definition}

Our discussion of pseudodifferential operators suggests the following
refinement of the notion of global hypoellipticity to the setting of
modulation spaces:

\begin{definition}
\label{defglobis}If $A:M_{1/v_{s}}^{\infty}(\mathbb{R}^{n})\longrightarrow
M_{1/v_{s}}^{\infty}(\mathbb{R}^{n})$ is a linear mapping, then it is
\textquotedblleft$s$-hypoelliptic\textquotedblright\ if we have%
\begin{equation}
\psi\in M_{1/v_{s_{0}}}^{\infty}(\mathbb{R}^{n})\text{ and }A\psi\in M_{v_{s}%
}^{1}(\mathbb{R}^{n})\Longrightarrow\psi\in M_{v_{s}}^{1}(\mathbb{R}^{n}).
\label{glob2}%
\end{equation}

\end{definition}

That this definition really provides us with a refinement of Definition
\ref{defglob} follows from the following observation:

\begin{lemma}
\label{lems}If $A$ is $s$-hypoelliptic for every $s\geq0$ then it is globally hypoelliptic.
\end{lemma}

\begin{proof}
Let $\psi\in\mathcal{S}^{\prime}(\mathbb{R}^{n})$; in view of the second
equality (\ref{30}) there exists $s_{0}$ such that $\psi\in M_{1/v_{s_{0}}%
}^{\infty}(\mathbb{R}^{n})$. The condition $A\psi\in M_{v_{s}}^{1}%
(\mathbb{R}^{n})$ for every $s$ then implies that $\psi\in M_{v_{s}}%
^{1}(\mathbb{R}^{n})$ for every $s\geq0\ $hence our claim in view of the first
equality (\ref{30}).
\end{proof}

Let us return to the Shubin classes we used in Section \ref{sec3} when we
studied spectral properties of Landau--Weyl operators. Using the properties of
these classes Shubin (\cite{sh87}, Chapter IV, \S 23) constructs a (left)
parametrix of $A$. i.e. a Weyl operator $B\in G\Gamma_{\rho}^{-m_{1},-m_{0}%
}(\mathbb{R}^{n})$ such that $BA=I+R$ where the kernel of $R$ is in
$\mathcal{S}(\mathbb{R}^{n}\times\mathbb{R}^{n})$; from the existence of such
a parametrix follows readily that:

\begin{proposition}
\label{proshu1}Any Weyl operator $A\in HG_{\rho}^{m_{1},m_{0}}(\mathbb{R}%
^{2n})$ is globally hypoelliptic.
\end{proposition}

Note that the previous proposition of Shubin remains true for the case of
modulation spaces, because all the arguments of his proof remain valid for
this more general class of function spaces.

In \cite{cogr06} Fredholm properties of (localization) pseudodifferential
operators on modulation spaces have been proved by Cordero and Gr\"{o}chenig.
These results provide natural generalizations of well-known results due to
Shubin on global hypoellipticity. We invoke their results to get some classes
of pseudodifferential operators that are $s$-hypoelliptic. We will use the
following refinement of Proposition \ref{proshu1}, also due to Shubin
(\cite{sh87}, Chapter IV, \S 25):

\begin{proposition}
\label{proshu2}$A\in HG_{\rho}^{m_{1},m_{0}}(\mathbb{R}^{2n})$ be such that
$\operatorname*{Ker}A=\operatorname*{Ker}A^{\ast}=\{0\}$. Then there exists
$B\in HG_{\rho}^{-m_{1},-m_{0}}(\mathbb{R}^{2n})$ such that $BA=AB=I$ (i.e.
$B$ is the inverse of $A$).
\end{proposition}

In \cite{cogr06} a class of symbol classes $\mathcal{M}_{v}$ is introduced,
which for the weight $v_{s}$ contains the classes $HG_{\rho}^{m_{1},m_{0}%
}(\mathbb{R}^{2n})$. Therefore by Theorem 7.1 in \cite{cogr06} we get a result
about the $s$-hypoellipticity for Landau--Weyl operators. The main result of
this section is the following global hypoellipticity result:

\begin{theorem}
\label{aglo}The Landau--Weyl\ operator $\widetilde{A}$ associated to an
operator $A\in G\Gamma_{\rho}^{m_{1},m_{0}}(\mathbb{R}^{2n})$ such that
$\operatorname*{Ker}A=\operatorname*{Ker}A^{\ast}=\{0\}$ is $s$-hypoelliptic
for each $s\geq0$ and hence also globally hypoelliptic.
\end{theorem}

\begin{proof}
In view of Proposition \ref{proshu2} the operator $A$ has an inverse $B$
belonging to $HG_{\rho}^{-m_{1},-m_{0}}(\mathbb{R}^{2n})$. In view of
Corollary \ref{corab} the LW operator $\widetilde{B}$ is then an inverse of
$\widetilde{A}$. Assume now that $\widetilde{A}\Psi=\Phi\in\mathcal{S}%
(\mathbb{R}^{2n})$; then $\Psi=\widetilde{B}\Phi$. The classes of symbols
studied in Theorem 7.1 and Corollary 7.2 in \cite{cogr06} contain the Shubin
classes $G\Gamma_{\rho}^{m_{1},m_{0}}(\mathbb{R}^{n})$ as one sees by an
elementary argument. Therefore it follows that any Weyl operator $A\in
H\Gamma_{\rho}^{m_{1},m_{0}}(\mathbb{R}^{2n})$ is $s$-hypoelliptic for every
$s\geq0$, and thus globally hypoelliptic in view of Lemma \ref{lems}.
\end{proof}

Consequently, the Landau--Weyl operator $\tilde{A}$ associated to an operator
$A\in G\Gamma_{\rho}^{m_{1},m_{0}}(\mathbb{R}^{n})$ such that
$\operatorname*{Ker}A=A^{\ast}=\{0\}$ is $s$-hypoelliptic for every $s\geq0$,
i.e. it is globally hypoelliptic.

\begin{remark}
The condition $A\in G\Gamma_{\rho}^{m_{1},m_{0}}(\mathbb{R}^{2n})$ does not
imply that $\widetilde{A}\in G\Gamma_{\rho}^{m_{1},m_{0}}(\mathbb{R}^{4n})$ as
is seen by inspection of formula (\ref{azzu}) for the symbol $\widetilde{a}$.
\end{remark}

Let us illustrate this when the symbol $a$ is a non-degenerate quadratic form:

\begin{example}
\label{corhyp}Let $a$ be a positive-definite quadratic form on $\mathbb{R}%
^{2n}$: $a(z)=\frac{1}{2}Mz\cdot z$ with $M=M^{T}>0$. Then $A$ is globally
hypoelliptic; in fact $a\in H\Gamma_{1}^{2,2}(\mathbb{R}^{2n})$ as is seen
using an adequate diagonalization of $M$.\ The operator $\widetilde{A}$ is
globally hypoelliptic. In particular the magnetic operator (\ref{magn1}) is
globally hypoelliptic.
\end{example}

Notice that in this example we have recovered the global hypoellipticity of
the magnetic operator obtained by Wong \cite{wo05-1} using very different
methods (the theory of special functions).

\subsection{Regularity results for the Schr\"{o}dinger equation}

Let us apply some of the previous results to the study of regularity
properties of the Schr\"{o}dinger equations%
\[
i\hbar\frac{\partial}{\partial t}\psi=H\psi\text{ \ , \ }i\hbar\frac{\partial
}{\partial t}\Psi=\widetilde{H}\Psi.
\]
Let the Hamiltonian function be a quadratic form:%
\[
H(z)=\tfrac{1}{2}Mz\cdot z\text{ \ , \ }M=M^{T}.
\]
The corresponding Hamiltonian flow consists of the linear symplectic mappings
$s_{t}=e^{tJM}$ and is hence a one-parameter subgroup of $\operatorname*{Sp}%
(2n,\mathbb{R})$. It follows from the theory of covering spaces that there is
a bijective correspondence between the one-parameter subgroups of the
symplectic group $\operatorname*{Sp}(2n,\mathbb{R})$ and those of the
metaplectic group $\operatorname*{Mp}(2n,\mathbb{R})$; let us denote this
correspondence by $\mu$. Thus
\[
\mu(s_{t})=S_{t}%
\]
means that if $(s_{t})$ is a one-parameter subgroup of $\operatorname*{Sp}%
(2n,\mathbb{R})$ then $(S_{t})$ is the only one-parameter subgroup of
$\operatorname*{Mp}(2n,\mathbb{R})$ whose projection is precisely $(s_{t})$.
We will similarly write%
\[
\widetilde{\mu}(s_{t})=\widetilde{S}_{t}%
\]
where $\widetilde{S}_{t}\in\operatorname*{Mp}(4n,\mathbb{R})$ is defined by
formula (\ref{mplw}).

The first part of following result is well-known:

\begin{proposition}
\label{metasch}Let $(s_{t})$ be the Hamiltonian flow determined by the
Hamilton equations $\dot{z}=J\partial_{z}H(z)=Mz$. The one parameter groups
$(S_{t})$ and $(\widetilde{S}_{t})$ defined by $S_{t}=\mu(s_{t})$ and
$\widetilde{S}_{t}=\widetilde{\mu}(s_{t})$ satisfy the Schr\"{o}dinger
equations%
\begin{equation}
i\hbar\frac{\partial}{\partial t}S_{t}=HS_{t}\text{ \ , \ }i\hbar
\frac{\partial}{\partial t}\widetilde{S}_{t}=\widetilde{H}\widetilde{S}_{t}
\label{sch}%
\end{equation}
where $H(x,-i\hbar\partial_{x})$ and $\widetilde{H}$ are the Weyl and LW
operators determined by the Hamiltonian function $H$.
\end{proposition}

\begin{proof}
That $S_{t}$ satisfies the first equation (\ref{sch}) is a classical result
(see for instance \cite{Birk,GS1,GS2}, for detailed accounts). That
$\widetilde{S}_{t}$ satisfies the second equation immediately follows.
\end{proof}

We next show that the spreading of wavefunction $\Psi$ and its evolution in
time can be controlled in terms of the spaces $L_{v_{s}}^{1}(\mathbb{R}^{2n})$.

\begin{proposition}
\label{schropsi1}Let $\Psi\in\mathcal{S}^{\prime}(\mathbb{R}^{2n})$ is a
solution of the Schr\"{o}dinger equation%
\begin{equation}
i\hbar\frac{\partial\Psi}{\partial t}=\widetilde{H}\Psi\text{ \ , \ }%
\Psi(\cdot,0)=\Psi_{0}. \label{schpsi}%
\end{equation}
If $\Psi_{0}\in\mathcal{H}_{\phi}\cap L_{v_{s}}^{1}(\mathbb{R}^{2n})$ for some
$\phi$ then $\Psi(\cdot,t)\in L_{v_{s}}^{1}(\mathbb{R}^{2n})$ for every
$t\in\mathbb{R}$.
\end{proposition}

\begin{proof}
Since $\Psi_{0}\in\mathcal{H}_{\phi}$ we have $\Psi_{0}=\mathcal{U}_{\phi}%
\psi_{0}$ for some $\psi_{0}\in L^{2}(\mathbb{R}^{n})$; the condition
$\Psi_{0}\in L_{v_{s}}^{1}(\mathbb{R}^{2n})$ implies that $\psi_{0}\in
M_{v_{s}}^{1}(\mathbb{R}^{n})$. Let $\psi$ be the unique solution of the
Cauchy problem%
\[
i\hbar\frac{\partial\psi}{\partial t}=H\psi\text{ , }\psi(\cdot,0)=\psi_{0};
\]
that solution is $\psi=S_{t}\psi_{0}$ in view of Proposition \ref{metasch},
hence $\psi(\cdot,t)\in M_{v_{s}}^{1}(\mathbb{R}^{n})$ for every
$t\in\mathbb{R}$. We claim that the (unique) solution of (\ref{schpsi}) with
$\Psi_{0}\in\mathcal{H}_{\phi}\cap L_{v_{s}}^{1}(\mathbb{R}^{2n})$ is
$\Psi=\mathcal{U}_{\phi}\psi$; the proposition will follows in view of the
definition of $M_{v_{s}}^{1}(\mathbb{R}^{n})$. Set $\Psi^{\prime}%
=\mathcal{U}_{\phi}\psi$. Since $\widetilde{H}\mathcal{U}_{\phi}%
=\mathcal{U}_{\phi}H(x,-i\hbar\partial_{x})$ in view of\ the second equality
(\ref{intertw}) in Proposition \ref{inter}, we have
\[
i\hbar\frac{\partial\Psi^{\prime}}{\partial t}=\mathcal{U}_{\phi}(i\hbar
\frac{\partial}{\partial t}\psi)=\widetilde{H}\mathcal{U}_{\phi}%
\psi=\widetilde{H}\Psi^{\prime}.
\]
Now $\Psi^{\prime}(\cdot,0)=\Psi_{0}$ hence $\Psi^{\prime}=\Psi$.
\end{proof}

\section{Generalization;\ Application to Deformation Quantization\label{sec5}}

\subsection{The operators $\widetilde{A}^{\gamma,\mu}$}

Let us show how to generalize the constructions above to the operators
$\widetilde{A}^{\gamma,\mu}$ corresponding to the more general quantization
rule (\ref{xjyj}).

We begin by noting that the operators $\widetilde{X}_{j}^{\gamma,\mu}$ and
$\widetilde{Y}_{j}^{\gamma,\mu}$ are obtained from $\widetilde{X}_{j}$ and
$\widetilde{Y}_{j}$ by conjugation with the metaplectic rescalings
$\widetilde{S}^{\gamma,\mu}$ defined in Lemma \ref{lemu}:
\begin{equation}
\widetilde{X}_{j}^{\gamma,\mu}=\widetilde{S}^{\gamma,\mu}\widetilde{X}%
_{j}(\widetilde{S}^{\gamma,\mu})^{-1}\text{ \ , \ }\widetilde{Y}_{j}%
^{\gamma,\mu}=\widetilde{S}^{\gamma,\mu}\widetilde{Y}_{j}(\widetilde
{S}^{\gamma,\mu})^{-1}.
\end{equation}
(the proof is purely computational and is therefore omitted). These formulae
suggest the following definition: for any LW operator $\widetilde{A}$ and
$(\gamma,\mu)\in\mathbb{R}^{2}$ such that $\gamma\mu\neq0$ we set
\begin{equation}
\widetilde{A}^{\gamma,\mu}=\widetilde{S}^{\gamma,\mu}\widetilde{A}%
(\widetilde{S}^{\gamma,\mu})^{-1}. \label{alu}%
\end{equation}
We have:

\begin{proposition}
(i) The contravariant symbol of $\widetilde{A}^{\gamma,\mu}:\mathcal{S}%
(\mathbb{R}^{2n})\longrightarrow\mathcal{S}^{\prime}(\mathbb{R}^{2n})$ is the
function%
\begin{equation}
\widetilde{a}^{\gamma,\mu}(x,y;p_{x},p_{y})=a(\tfrac{\gamma}{2}x-\tfrac{1}%
{\mu}p_{y},\tfrac{\mu}{2}y+\tfrac{1}{\gamma}p_{x}) \label{asymbul}%
\end{equation}
where $a$ is the contravariant symbol of $A$. (ii) We have
\begin{equation}
\widetilde{A}^{\gamma,\mu}=\left(  \tfrac{1}{2\pi\hbar}\right)  ^{n}%
\int_{\mathbb{R}^{2n}}a_{\sigma}^{\gamma,\mu}(z)\widetilde{T}^{\gamma,\mu
}(z)dz \label{altaul}%
\end{equation}
(Bochner integral) where $a_{\sigma}^{\gamma,\mu}=\widetilde{S}^{\gamma,\mu
}a_{\sigma}$ and $\widetilde{T}^{\gamma,\mu}(z)$ is the unitary operator
defined by%
\begin{equation}
\widetilde{T}^{\gamma,\mu}(z_{0})\Psi(z)=e^{-\frac{i\gamma\mu}{2\hbar}%
\sigma(z,z_{0})}\Psi(z-z_{0}). \label{numero}%
\end{equation}

\end{proposition}

\begin{proof}
Formula (\ref{asymbul}) follows from the symplectic covariance (\ref{metaco})
of Weyl calculus taking (\ref{sdiag}) into account. Formula (\ref{altaul})
follows, by a change of variables in definition (\ref{ahwbis}) of
$\widetilde{A}$.
\end{proof}

The following intertwining result is a straightforward consequence of
Proposition \ref{inter}:

\begin{corollary}
(i) The mapping $\mathcal{U}_{\phi}^{\gamma,\mu}=\widetilde{S}^{\gamma,\mu
}\mathcal{U}_{\phi}$ is an isometry of $L^{2}(\mathbb{R}^{n})$ onto the closed
subspace $\mathcal{H}_{\phi}^{\gamma,\mu}=\widetilde{S}^{\gamma,\mu
}\mathcal{H}_{\phi}$ of $L^{2}(\mathbb{R}^{2n})$; explicitly
\begin{equation}
\mathcal{U}_{\phi}^{\gamma,\mu}\psi(z)=\left(  \tfrac{\pi\gamma\mu\hbar}%
{2}\right)  ^{n/2}W(\psi,\phi)(\tfrac{1}{2}\gamma x,\tfrac{1}{2}\mu y).
\label{uexplicit}%
\end{equation}
(ii) The operator $\widetilde{A}^{\gamma,\mu}$ satisfies the intertwining
formula%
\begin{equation}
\text{\ }\widetilde{A}^{\gamma,\mu}\mathcal{U}_{\phi}^{\gamma,\mu}%
=\mathcal{U}_{\phi}^{\gamma,\mu}\widehat{A}\text{\ \ with }\mathcal{U}_{\phi
}^{\gamma,\mu}=\widetilde{S}^{\gamma,\mu}\mathcal{U}_{\phi}. \label{aulm}%
\end{equation}

\end{corollary}

\begin{proof}
(i) $\mathcal{U}_{\phi}^{\gamma,\mu}$ is the compose of two isometries hence
an isometry. $\mathcal{H}_{\phi}^{\gamma,\mu}$ is closed because
$\mathcal{H}_{\phi}$ is, and $\widetilde{S}^{\gamma,\mu}$ is an isomorphism
$L^{2}(\mathbb{R}^{n})\longrightarrow L^{2}(\mathbb{R}^{n})$. (ii) Formula
(\ref{aulm}) immediately follows from the definitions of \ $\widetilde
{A}^{\gamma,\mu}$ and $\mathcal{U}_{\phi}^{\gamma,\mu}$ and the second
intertwining formula (\ref{intertw}).
\end{proof}

\subsection{The Moyal product and Deformation Quantization}

The Moyal product plays a central role in deformation quantization of Flato
and Sternheimer \cite{BFFLS}. Let $H$ be a Hamiltonian function and assume
that $\Psi\in\mathcal{S}(\mathbb{R}^{n})$; the Moyal product \cite{BFFLS}
$H\star_{\hbar}\Psi$ is defined by
\begin{equation}
(H\star_{\hbar}\Psi)(z)=\left(  \tfrac{1}{4\pi\hbar}\right)  ^{2n}%
\iint\nolimits_{\mathbb{R}^{n}\times\mathbb{R}^{n}}e^{\frac{i}{2\hbar}%
\sigma(u,v)}H(z+\tfrac{1}{2}u)\Psi(z-\tfrac{1}{2}v)dudv; \label{cz}%
\end{equation}
when $\hbar=1/2\pi$ it reduces to the twisted product $\#$ familiar from
standard Weyl calculus: $H\star_{1/2\pi}\Psi=H\#\Psi$.

We claim that%
\begin{equation}
H\star_{\hbar}\Psi=\widetilde{H}^{2,1}\Psi=H(x-\tfrac{1}{2}i\hbar\partial
_{p},p+\tfrac{1}{2}i\hbar\partial_{x})\Psi. \label{hstarpsi}%
\end{equation}
The proof is similar to that of Theorem \ref{proplww}. Let us view
$\Psi\longmapsto H\star_{\hbar}\Psi$ as a Weyl operator, denoted by
$H\star_{\hbar}$. Using formula (\ref{cz}) the distributional kernel of
$H\star_{\hbar}$ is given by%
\begin{equation}
\mathcal{K}_{H\star_{\hbar}}(z,y)=\left(  \tfrac{1}{2\pi\hbar}\right)
^{2n}\int_{\mathbb{R}^{2n}}e^{\frac{i}{\hbar}\sigma(u,z-y)}H(z-\tfrac{1}%
{2}u)du \label{k}%
\end{equation}
hence, using (\ref{axy}) and the Fourier inversion formula, the contravariant
symbol of $H\star_{\hbar}$ is given by%
\[
\mathbb{H}(z,\zeta)=\int_{\mathbb{R}^{2n}}e^{-\frac{i}{\hbar}\zeta\cdot\eta
}\mathcal{K}_{H\star_{\hbar}}(z+\tfrac{1}{2}\eta,z-\tfrac{1}{2}\eta)d\eta.
\]
Using (\ref{k}) and performing the change of variables $u=2z+\eta-z^{\prime}$
we get
\[
\mathcal{K}_{H\star_{\hbar}}(z+\tfrac{1}{2}\eta,z-\tfrac{1}{2}\eta
)d\eta=\left(  \tfrac{1}{2\pi\hbar}\right)  ^{2n}e^{\frac{2i}{\hbar}%
\sigma(z,\eta)}\int_{\mathbb{R}^{2n}}e^{\frac{i}{\hbar}\sigma(\eta,z^{\prime
})}H(\tfrac{1}{2}z^{\prime})dz^{\prime};
\]
setting $H(\tfrac{1}{2}z^{\prime})=H_{1/2}(z^{\prime})$ the integral is
$\left(  2\pi\hbar\right)  ^{n}$ times the symplectic Fourier transform
$F_{\sigma}^{\hbar}H_{1/2}(-\eta)=(H_{1/2})_{\sigma}(-\eta)$ so that%
\begin{align*}
\mathbb{H}(\tfrac{1}{2}z,\zeta)  &  =\left(  \tfrac{1}{2\pi\hbar}\right)
^{n}\int_{\mathbb{R}^{2n}}e^{-\frac{i}{\hbar}\zeta\cdot\eta}e^{\frac{i}{\hbar
}\sigma(z,\eta)}(H_{1/2})_{\sigma}(-\eta)d\eta\\
&  =\left(  \tfrac{1}{2\pi\hbar}\right)  ^{n}\int_{\mathbb{R}^{2n}}%
e^{-\frac{i}{\hbar}\sigma(z+J\zeta,\eta)}(H_{1/2})_{\sigma}(\eta)d\eta
\end{align*}
Since the second equality is the inverse symplectic Fourier transform of
$(H_{1/2})_{\sigma}$ calculated at $z+J\zeta$ we finally get%
\begin{equation}
\mathbb{H}(z,\zeta)=H(x+\tfrac{1}{2}\zeta_{p},p-\tfrac{1}{2}\zeta_{x})
\label{hsymb}%
\end{equation}
with $\zeta=(\zeta_{x},\zeta_{p})$.

An immediate consequence of these results is:

\begin{proposition}
The isometries $\mathcal{U}_{\phi}^{2,1}$ defined by the formula%
\begin{equation}
\mathcal{U}_{\phi}^{2,1}\psi(z)=\left(  \pi\hbar\right)  ^{n/2}W(\psi,\phi)(z)
\label{u21}%
\end{equation}
satisfy the intertwining relation
\[
(H\star_{\hbar}\Psi)\mathcal{U}_{\phi}^{2,1}=\mathcal{U}_{\phi}^{2,1}%
\widehat{H}.
\]

\end{proposition}

\begin{proof}
Formula (\ref{u21}) is just (\ref{uexplicit}) with $\gamma=2,\mu=1.$
\end{proof}

\section{Concluding Remarks}

Due to limitations of length and time we have only been able to give a few
applications of the theory of modulation spaces to the Landau--Weyl calculus.
Modulation spaces and related topics shave turned out to be the proper setting
for the discussion of pseudodifferential operators in the last decade, see for
instance the papers \cite{grhe99,grhe04} and the references therein; for
related topics such as the spaces $\ell^{q}(L^{p})$ see for instance the work
of Birman and Solomjak \cite{BS}, Christ and Kiselev \cite{CK} or Simon
\cite{si79}. Recently modulation spaces have also found various applications
in the study of Schr\"{o}dinger operators (see Cordero and Nicola
\cite{coni06,coni08}).

\end{document}